\documentclass[lettersize,journal]{IEEEtran}
\usepackage{amsmath,amsfonts}
\usepackage{algorithmic}
\usepackage{array}
\usepackage[caption=false,font=normalsize,labelfont=sf,textfont=sf]{subfig}
\usepackage{textcomp}
\usepackage{stfloats}
\usepackage{url}
\usepackage{verbatim}
\usepackage{graphicx}
\usepackage[table]{xcolor} 
\def\BibTeX{{\rm B\kern-.05em{\sc i\kern-.025em b}\kern-.08em
    T\kern-.1667em\lower.7ex\hbox{E}\kern-.125emX}}
\usepackage{balance}

\usepackage{setspace}
\usepackage{subfloat}
\usepackage{mathtools}
\usepackage{amsmath,amsfonts,amssymb,amsthm,mathrsfs, bm}
\usepackage[utf8]{inputenc}
\usepackage[swedish, english]{babel}
\usepackage{csquotes}
\usepackage{subfig}
\usepackage{graphicx}
\usepackage{supertabular}                                             
\usepackage{booktabs} 

\usepackage{graphicx}
\usepackage{epstopdf}
\usepackage{microtype}
\usepackage{multirow}
\usepackage{hyperref}
\usepackage{xcolor}
\usepackage{authblk}
\usepackage{datetime2}	
\usepackage{setspace}	
\theoremstyle{plain}
\newtheorem{theorem}{Theorem}[section]
\newtheorem{lemma}{Lemma}[section]

\newtheorem{proposition}{Proposition}[section]

\usepackage{amsmath,amsfonts,amssymb,amsthm,mathrsfs, bm}
\usepackage{stmaryrd}
\usepackage{mathtools, cuted}
\usepackage{graphicx}
\usepackage{tikz}
\usepackage[utf8]{inputenc}
\usepackage{pgfplots} 
\usepackage{pgfgantt}
\usepackage{pdflscape}
\pgfplotsset{compat=newest} 
\pgfplotsset{plot coordinates/math parser=false}

\usepackage{endnotes}
\usepackage{commath}
\usepackage{gensymb}
\usepackage{mathtools}
\usepackage{tikz} 
\usetikzlibrary{calc,quotes,angles}
\usepackage{tkz-euclide}
\usetikzlibrary{automata,positioning} 

\usepackage{xcolor}
\usepackage{tikz,tkz-euclide,siunitx}
\usepackage{etoolbox} 

\usetikzlibrary{quotes,arrows.meta,angles,calc,shadings,positioning,babel}

\newtheorem{problem}{Problem}[section]
\newtheorem{remark}{Remark}[section]

\usepackage{endnotes}
\usepackage{commath}
\usepackage{gensymb}
\usepackage{textgreek} 

\makeatletter
\patchcmd{\tkz@DrawLine}{\begingroup}{\begingroup\makeatletter}{}{}
\makeatother

\allowdisplaybreaks
\DeclareMathOperator{\diag}{\mathrm{diag}}

\DeclareMathOperator{\esup}{ess\, sup}

\makeatletter
\newcommand\makebig[2]{%
  \@xp\newcommand\@xp*\csname#1\endcsname{\bBigg@{#2}}%
  \@xp\newcommand\@xp*\csname#1l\endcsname{\@xp\mathopen\csname#1\endcsname}%
  \@xp\newcommand\@xp*\csname#1r\endcsname{\@xp\mathclose\csname#1\endcsname}%
}
\makeatother

\providecommand*{\ped}[1]{%
\ensuremath{_\textnormal{#1}}}
\providecommand*{\ap}[1]{%
\ensuremath{^\textnormal{#1}}}
\providecommand*{\eu}%
{\ensuremath{\mathrm{e}}}
\providecommand*{\im}%
{\ensuremath{\mathrm{i}}}
\providecommand*{\GammaF}%
{\ensuremath{\mathrm{\Gamma}}}
\providecommand*{\BetaF}%
{\ensuremath{\mathrm{\Beta}}}

\makebig{biggg} {3.0}
\makebig{Biggg} {3.5}
\makebig{bigggg}{4.0}
\makebig{Bigggg}{4.5}
\makebig{biggggg}{5.0}
\makebig{Biggggg}{5.5}
\usepackage{longtable}
\usepackage{xcolor}

\usepackage{hyperref}
\hypersetup{
    colorlinks = true,
    linkbordercolor = {white},
            linkcolor=blue,
            bookmarks=true,
            filecolor=blue,
            urlcolor=blue,
            citecolor=blue
}

\begin{document}
\title{Lateral tracking control of all-wheel \\steering vehicles with intelligent tires}
\author{IEEE Publication Technology Department
\thanks{Manuscript created October, 2020; This work was developed by the IEEE Publication Technology Department. This work is distributed under the \LaTeX \ Project Public License (LPPL) ( http://www.latex-project.org/ ) version 1.3. A copy of the LPPL, version 1.3, is included in the base \LaTeX \ documentation of all distributions of \LaTeX \ released 2003/12/01 or later. The opinions expressed here are entirely that of the author. No warranty is expressed or implied. User assumes all risk.}}

\author{Luigi Romano~\IEEEmembership{Member,~IEEE}, Ole Morten Aamo~\IEEEmembership{Senior Member,~IEEE}, Jan Åslund, Erik Frisk
\thanks{L. Romano is with the Department of Electrical Engineering, Linköping University, Linköping, Sweden, and the Department of Engineering Cybernetics, NTNU, Trondheim, Norway.}
\thanks{O. M. Aamo is with the Department of Engineering Cybernetics, NTNU, Trondheim, Norway.}
\thanks{J. Åslund and E. Frisk are with the Department of Electrical Engineering, Linköping University, Linköping, Sweden.}
}

\markboth{Journal of \LaTeX\ Class Files, June~2024}%
{How to Use the IEEEtran \LaTeX \ Templates}

\maketitle

\begin{abstract}
The accurate characterization of tire dynamics is critical for advancing control strategies in autonomous road vehicles, as tire behavior significantly influences handling and stability through the generation of forces and moments at the tire-road interface. Smart tire technologies have emerged as a promising tool for sensing key variables such as road friction, tire pressure, and wear states, and for estimating kinematic and dynamic states like vehicle speed and tire forces. However, most existing estimation and control algorithms rely on empirical correlations or machine learning approaches, which require extensive calibration and can be sensitive to variations in operating conditions. In contrast, model-based techniques, which leverage infinite-dimensional representations of tire dynamics using \emph{partial differential equations} (PDEs), offer a more robust approach. This paper proposes a novel model-based, output-feedback lateral tracking control strategy for all-wheel steering vehicles that integrates distributed tire dynamics with smart tire technologies. The primary contributions include the suppression of \emph{micro-shimmy} phenomena at low speeds and path-following via force control, achieved through the estimation of tire slip angles, vehicle kinematics, and lateral tire forces. The proposed controller and observer are based on formulations using ODE-PDE systems, representing rigid body dynamics and distributed tire behavior. This work marks the first rigorous control strategy for vehicular systems equipped with distributed tire representations in conjunction with smart tire technologies.
\end{abstract}


\section{Introduction}\label{sect:Intro}
\IEEEPARstart{T}{he} development of sustainable, intelligent, and safer transportation systems necessitates the implementation of advanced control strategies that support autonomous navigation and driving. In the domain of road vehicles, an accurate characterization of tire behavior, which governs the generation of forces and moments at the tire-road interface, is critical for the design of control and estimation algorithms aimed at enhancing vehicular performance \cite{Gerdes1,Gerdes2,Gerdes3,JazarNew,IEEEVT1,IEEEVT2}. In fact, tires exhibit a complex dynamical behavior that may introduce significant delay effects – also referred to as as \textit{relaxation} phenomena – in the process of generation of forces and moments \cite{Pacejka2,LibroMio,CarcassDyn}, with important implications on handling performance and stability \cite{Guiggiani}. Specifically, two main factors may be identified that contribute, to a different extent, to the transient response of the tire subjected to translational slip and spin inputs. The first one is connected to the local deformation of the tread rubber that continuously enters and leaves the contact patch during the rolling of the wheel \cite{LibroMio,CarcassDyn,Guiggiani,Meccanica2,LuGreSpin}. This rubber flow is responsible for exciting delay dynamics that are well described by transport-like equations, possibly incorporating nonlinear or boundary terms. The second unsteady effect, of a global nature, may instead be ascribed to the compliance of the carcass element and, when the tire operates in its linear region, introduces delay responses that are similar to those of a first-order system \cite{Pacejka2,Guiggiani}. Recent analyses seem to confirm that both phenomena affect crucially the longitudinal and lateral dynamics of road vehicles \cite{Guiggiani}, as well as the energy consumption originating from slip losses. In the context outlined above, the necessity of accurately estimating tire forces is self-evident if not pleonastic \cite{Rajamani,Nielsen,Savaresi,IEEEVT3}. Despite their critical importance, in the scientific literature, model-based detection of tire forces and slip angles has largely been approached using either static models or lumped formulations that disregard the local deformation of rubber deformation inside the contact patch \cite{RajamaniCC,Sideslip1,Sideslip2,Doumiati2,Hsu}. These simplifications are often justified by the limited availability of measurable signals, typically restricted to kinematic variables such as yaw rate, vehicle acceleration, and possibly aligning torque \cite{Shao1,Shao3,AdaptiveCornering}.

In recent years, however, smart or intelligent tire technologies have garnered increasing attention due to their superior capability in sensing critical information such as friction and road conditions, outperforming the standard instrumentation available in passenger vehicles \cite{IntTireSyst,IT1,IT2}.
Indeed, smart tire sensors enable the estimation of several key quantities, including contact patch length, road grip, tire pressure, and wear states \cite{ITTerrain,ITwear}. Apart from allowing to actively monitor tire conditions, intelligent tire technologies provide several benefits also concerning the estimation of kinematic variables like vehicle's speed, yaw rate, and sideslip angle, as well as dynamic states such as tire forces and moments \cite{ITSideSlip,Benefits,Erdogan}. The majority of estimation and control algorithms presented in the literature are predominantly based on empirical correlations between tire operating conditions and signal features detected by strain sensors or accelerometers mounted on the inner liner of the tire. Some methodologies integrate traditional techniques with machine learning algorithms, which can infer friction levels and the forces generated at the tire-road contact patch \cite{IntTireSyst,IT3}. The main limitation of these approaches is that they require extensive calibration and training; moreover, the accurate estimation of tire features might be jeopardized by minimal variations in the operating conditions. In contrast, only a limited number of model-based estimation techniques utilizing intelligent tire technologies have been proposed. These approaches often involve infinite-dimensional representations of tire dynamics, typically formulated using \emph{partial differential equations} (PDEs). Amongst the most commonly employed models in this context are the brush models \cite{Erdogan,ModelBIT1,ModelBIT3}, and the flexible ring model \cite{ModelBIT2,Unina1,Unina3,Unina2}.

In some circumstances, the use of PDEs in the modeling of tire behavior may also be essential for the direct synthesis of control algorithms. Recent studies indicate that unstable vehicle dynamics, triggered by the distributed nature of tire deformation within the contact patch, can manifest even at relatively low speeds. Such dynamics is associated with the so-called \emph{micro-shimmy} phenomenon, which contributes to increased energy consumption and slip losses \cite{Takacs1,Takacs2,Takacs3,Takacs4,Takacs5,BicyclePDE}. This behavior, not captured by conventional lumped models formulated using \emph{ordinary differential equations} (ODEs), demands the adoption of more sophisticated infinite-dimensional representations of tire and vehicle dynamics. 

This paper presents the theoretical development of a model-based, output-feedback lateral tracking control strategy for all-wheel steering vehicles, considering explicitly distributed tire dynamics in conjunction with smart tire technologies. Specifically, the paper addresses two main challenges: the suppression of micro-shimmy phenomena at low speeds and trajectory tracking via force control. The proposed controller relies on the detection of the vehicle's kinematic variables, tire slip angles, and lateral forces. Therefore, the estimation problem is also addressed via observer design, assuming that smart tire sensors are available in addition to standard signals. Since both the mitigation of micro-shimmy oscillations and the adoption of smart tire technologies demand the adoption of distributed tire representations, the models employed in this study are formulated as ODE-PDE systems, where the ODEs describe the rigid body vehicle dynamics, whilst the PDEs model the distributed tire behavior. To the best of the authors' knowledge, this work represents the first rigorous attempt at controlling vehicular systems equipped with a transient, distributed tire representation, when accounting for smart tire technologies. 

Due to the inherent complexity of the model employed in this study, some simplifying assumptions have been introduced. First, as previously noted, the analysis focuses on all-wheel steering vehicles. Whilst this may initially seem like a significant constraint, it is anticipated that future cars will increasingly adopt such steering systems \cite{ChinaFeng,LarsAVEC,Jerrelind}, facilitated by the advancement and integration of steer-by-wire technologies and/or autonomous driving \cite{AllWheel1,AllWheel2,SteerByWireUlsoy}. Additionally, it is assumed that data from smart tire devices are collected continuously over time. This assumption implies a need for a larger number of sensors than what is typically equipped by passenger cars, or, alternatively, the use of more costly optical devices \cite{Tuononen1,Tuononen2}. Nevertheless, the cost of smart tire sensors is expected to substantially decrease in the coming years. In any case, the two major assumptions introduced in this paper may provide important indications to manufacturers for the design of next-gen vehicles.

The remainder of the manuscript is organized as follows. Section \ref{sect:Model} introduces the model adopted in the paper, discusses the main assumptions behind the proposed formulation, and recollects some preliminary results about its stability properties. Moreover, the main objective of the work is also formalized. Then, Section \ref{sect:StateFeed} presents the state-feedback control strategy. The state observer and the corresponding output-feedback controller are then developed in Section \ref{sect:observerMain}. Finally, the performance of the observer and controller is tested in Section \ref{sect:Val} considering different scenarios and models with degree of fidelity. The main conclusions, along with some possible directions for future research, are finally reported in Section \ref{sect:concl}.

Before moving to the core of the manuscript, it is worth clarifying that the perspective of this paper is primarily theoretical, albeit key implementational aspects of the proposed algorithms are also discussed, with a particular focus on the smart tire sensors (and related signals) required for observer design. Eventually, the insights derived from the present study have the potential to catalyze new research directions in the development of advanced mechatronic solutions for vehicular systems equipped with intelligent tire technologies.

\subsection{Notation}\label{sect:Notation}
In this paper, $\mathbb{Z}$ and $\mathbb{R}$ denote the set of rational and real numbers; $\mathbb{R}_{>0}$ and $\mathbb{R}_{\geq 0}$ indicate the set of positive real numbers excluding and including zero, respectively. Similarly, $\mathbb{C}$ is the set of complex numbers; $\mathbb{C}_{>0}$ and $\mathbb{C}_{\geq 0}$ denote the sets of all complex numbers whose real part is larger than and larger than or equal to zero, respectively. The group of $n\times m$ matrices with values in a field $\mathbb{F}$ is denoted by $\mathbf{M}_{n\times m}(\mathbb{F})$ (abbreviated as $\mathbf{M}_{n}(\mathbb{F})$ whenever $m=n$). For $\mathbb{F} = \mathbb{R}$, $\mathbb{R}_{>0}$, or $\mathbb{R}_{\geq0}$, $\mathbf{GL}_n(\mathbb{F})$ represents the group of invertible matrices with values in $\mathbb{F}$; the identity matrix on $\mathbb{R}^n$ is indicated with $\mathbf{I}_n$.
The standard Euclidean norm on $\mathbb{R}^n$ is indicated with $\norm{\cdot}_2$; matrix norms are simply denoted by $\norm{\cdot}$.
$L^2((0,1);\mathbb{R}^n)$ denotes the Hilbert space of square-integrable functions on $(0,1)$ with values in $\mathbb{R}^n$, and is equipped with norm $\norm{\bm{v}(\cdot)}_{L^2((0,1);\mathbb{R}^n)}^2 = \int_0^1 \norm{\bm{v}(\xi)}_2^2 \dif \xi$. $H^1((0,1);\mathbb{R}^n)$ indicates the Sobolev space of functions $\bm{v} \in L^2((0,1);\mathbb{R}^n)$ whose weak derivative also belongs to $L^2((0,1);\mathbb{R}^n)$.
Given a generic Hilbert space $\mathcal{V}$, $C^0([0,T];\mathcal{V})$, $L^1((0,T);\mathcal{V})$, $L^2((0,T);\mathcal{V})$, and $L^\infty([0,T];\mathcal{V})$ denote respectively the spaces of continuous, integrable, square-integrable, and (essentially) bounded functions on $[0,T]$ with values in $\mathcal{V}$ (for $T = \infty$, the interval $[0,T]$ is identified with $\mathbb{R}_{\geq 0}$). The following notation is adapted from \cite{Zwarth,CurtainAutomatica}.
Consider $C^\omega(\mathbb{C}_{>0};\mathbf{M}_{n\times m}(\mathbb{C}))$, the set of all analytic functions from $\mathbb{C}_{>0}$ to $\mathbb{C}^{n\times m}$; the Hardy space $H^\infty(\mathbb{C}_{>0};\mathbf{M}_{n\times m}(\mathbb{C}))$ is defined as $H^\infty(\mathbb{C}_{>0};\mathbf{M}_{n\times m}(\mathbb{C})) \triangleq \{ \mathbf{G} \in C^\omega(\mathbb{C}_{>0};\mathbf{M}_{n\times m}(\mathbb{C})) \mathrel{|} \norm{\mathbf{G}(\cdot)}_\infty < \infty\}$, with $\norm{\mathbf{G}(\cdot)}_\infty \triangleq \esup_{ s\in \mathbb{C}_{>0}} \norm{\mathbf{G}(s)}$. Finally, the Laplace transform of a variable $\bm{v}(t) \in \mathcal{V}$ is denoted by $\widehat{\bm{v}}(s) = (\mathcal{L}\bm{v})(s)$.

\section{Model description and problem formulation}\label{sect:Model}
The present Section details the main equations governing the cornering dynamics of a single-track model with distributed tires. The formulation adopted in the paper has been introduced by the authors in \cite{BicyclePDE}. The stability of the model, along with that of some of its subsystems, is briefly discussed in the frequency domain, based on the results derived in \cite{BicyclePDE}.

\subsection{Model description}\label{sect:MDesc}
\begin{figure}
\centering
\includegraphics[width=0.9\linewidth]{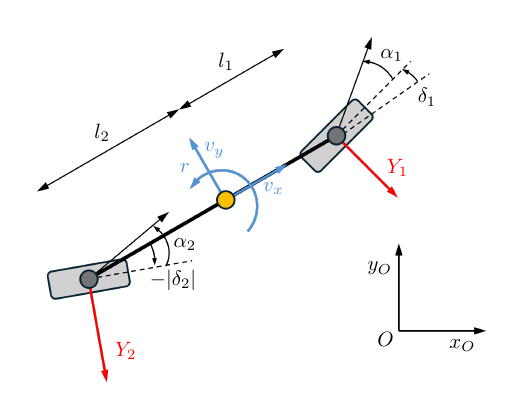} 
\caption{Single-track vehicle model with four-wheel steering. The kinematic variables are depicted in blue, whereas the dynamic ones in red.}
\label{figureForcePostdoc}
\end{figure}

\begin{figure}
\centering
\includegraphics[width=1\linewidth]{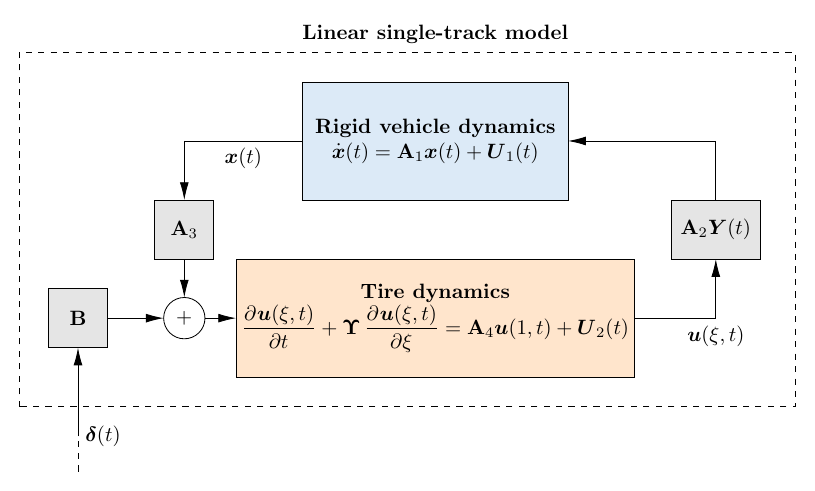} 
\caption{Schematic representation of the ODE-PDE interconnection \eqref{eq:ssOriginal}-\eqref{eq:B1Origin}, with $\mathbb{R}^2\ni \bm{U}_1(t) \triangleq \mathbf{A}_2\bm{Y}(t)$, $\mathbb{R}^2 \ni \bm{U}_2(t) \triangleq \mathbf{A}_3\bm{x}(t) + \mathbf{B}\bm{\delta}(t)$, and $\bm{Y}(t)$ given according to \eqref{eq:tireForce}.}
\label{figure:system0}
\end{figure}

Consider the ODE system governing the rigid dynamics of a linear single-track model, as illustrated in Figure \ref{figureForcePostdoc}:
\begin{subequations}\label{eq:bycicleVer1}
\begin{align}
 m\dot{v}_y(t)  &= -Y_{1}(t) - Y_{2}(t)-mv_x r(t), &&  \label{eq:vy}\\
I_z\dot{r}(t) & = -l_1Y_{1}(t)  +l_2Y_{2}(t), &&  t \in (0,T),  \label{eq:r}
\end{align}
\end{subequations}
where the state vector $[v_y(t) \; r(t)]^{\mathrm{T}} \in \mathbb{R}^2$ contains the vehicle's lateral velocity and yaw rate, $v_x \in \mathbb{R}_{>0}$ denotes its constant longitudinal speed, $m\in \mathbb{R}_{>0}$ is the vehicle's mass, $I_z\in \mathbb{R}_{>0}$ its inertia around the vertical axis, $l_1,l_2\in \mathbb{R}_{>0}$ represent the axle lengths, and $Y_{i}(t) \in \mathbb{R}$, $i \in \{1,2\}$, are the axle lateral forces, collected into the vector $\mathbb{R}^2 \ni \bm{Y}(t) =[Y_1(t)\;Y_2(t)]^{\mathrm{T}}$. For small steering angles $\delta_1(t), \delta_2(t) \in \mathbb{R}$ at the front and rear axle, these may be modeled as
\begin{align}\label{eq:tireForce}
\bm{Y}(t) = \begin{bmatrix} \dfrac{C_1}{2a_1} & 0 \\ 0 & \dfrac{C_2}{2a_2}\end{bmatrix}\int_{0}^1 \bm{u}(\xi,t) \dif \xi, && t \in [0,T).
\end{align}
The distributed state vector $\mathbb{R}^2 \ni \bm{u}(\xi,t) = [u_{1}(\xi,t) \; u_{2}(\xi,t)]^{\mathrm{T}}$ in \eqref{eq:tireForce} collects the local deformation of the front and rear tires inside the contact patches\footnote{More precisely, $u_1(\xi,t)$ represents the sum of the bristle's lateral deflection inside the front left and right tires, whereas $u_2(\xi,t)$ is the sum of the lateral deflections of rear left and right tires.}, and $a_i, C_i \in \mathbb{R}_{>0}$, $i \in \{1,2\}$, denote the front and rear tire's contact patch semilength and the cornering stiffness of the axles, respectively. In turn, the dynamics of $\bm{u}(\xi,t)$ obeys
\begin{subequations}\label{eq:PDEsfullcontST}
\begin{align}
\begin{split}
\dpd{u_{1}(\xi,t)}{t} & + \dfrac{v_x}{2a_1} \dpd{u_{1}(\xi,t)}{\xi}  =v_x \dfrac{2a_1}{\lambda_1}\alpha_1\bigl(v_y(t), r(t),\delta_1(t) \bigr) \\
& \quad + v_x\dfrac {C_1}{2a_1\lambda_1 w_1} u_{1}(1,t),   \label{eq:PDEVer1} 
\end{split}\\
\begin{split}
\dpd{u_{2}(\xi,t)}{t} & + \dfrac{v_x}{2a_2} \dpd{u_{2}(\xi,t)}{\xi} = v_x\dfrac{2a_2}{\lambda_2}\alpha_2\bigl(v_y(t), r(t),\delta_2(t)\bigr) \\
& \quad + v_x\dfrac{C_2}{2a_2\lambda_2 w_2} u_{2}(1,t), \quad (x,t) \in (0, 1)\times (0,T). \label{eq:PDE2Ver2}
\end{split}
\end{align}
\end{subequations}
in which the structural parameters $w_i, \lambda_i \in \mathbb{R}_{>0}$, $i \in \{1,2\}$, denote the carcass stiffnesses and relaxation lengths of the front and rear axles, respectively. In particular, according to \cite{LibroMio} (Chapter 6) the relaxation lengths are given by
\begin{align}\label{eq:relaxLength}
\lambda_i \triangleq a_i + C_i/w_i, && i \in \{1,2\}.
\end{align}
Finally, owing to the assumption of small steering inputs, the slip angles in \eqref{eq:PDEsfullcontST} read
\begin{subequations}\label{eq:lateralSlips}
\begin{align}
\alpha_1(v_y, r,\delta_1 ) & = \dfrac{v_y+l_1r}{v_x}-\delta_1, \\
\alpha_2(v_y,r,\delta_2) & = \dfrac{v_y-l_2r}{v_x}-\delta_2.
\end{align}
\end{subequations}
The PDEs \eqref{eq:PDEsfullcontST} come equipped with the following set of \emph{boundary} (BCs) and \emph{initial conditions} (ICs):
\begin{subequations}\label{eq:IBCs}
\begin{align}
& \text{BC:} && \bm{u}(0,t) =\bm{0},&&t\in (0,T), \label{eq:BCs}\\
& \text{IC:} && \bm{u}(\xi,0) = \bm{u}_{0}(\xi), && \xi \in (0,1).
\end{align}
\end{subequations}
Equations \eqref{eq:bycicleVer1}-\eqref{eq:BCs} may be recast in state-space form as
\begin{subequations}\label{eq:ssOriginal}
\begin{align}
& \dot{\bm{x}}(t) = \mathbf{A}_1\bm{x}(t) + \mathbf{A}_2 \bm{Y}(t), \quad t \in (0,T), \label{eq:ssOriginalODE}\\
\begin{split}
& \dpd{\bm{u}(\xi,t)}{t} + \mathbf{\Upsilon} \dpd{\bm{u}(\xi,t)}{\xi} = \mathbf{A}_3\bm{x}(t) + \mathbf{A}_4\bm{u}(1,t) + \mathbf{B}\bm{\delta}(t), \\
& \qquad \qquad \qquad \qquad \qquad \quad (\xi,t) \in (0,1) \times (0,T), 
\end{split}\label{eq:ssOriginalPDE} \\
& \bm{u}(0,t) = \bm{0}, \quad t\in(0,T), \label{eq:ssOriginalBC}
\end{align}
\end{subequations}
where $\mathbb{R}^2 \ni \bm{\delta}(t) = [\delta_1(t) \; \delta_2(t)]^{\mathrm{T}}$ denotes the input vector, the matrix $\mathbf{GL}_2(\mathbb{R}) \ni \mathbf{\Upsilon} = \diag\{\upsilon_1, \upsilon_2\}$ collects the transport velocities, i.e.,
\begin{align}\label{eq:Upsilon}
\mathbf{\Upsilon} = \begin{bmatrix} \upsilon_1 & 0 \\ 0 & \upsilon_2\end{bmatrix} \triangleq \begin{bmatrix}\dfrac{v_x}{2a_1} & 0 \\  0 & \dfrac{v_x}{2a_2}\end{bmatrix},
\end{align}
the matrices $\mathbf{A}_1, \mathbf{A}_2, \mathbf{A}_3, \mathbf{A}_4 \in \mathbf{M}_2(\mathbb{R})$ read
\begin{align}\label{eq:A1s}
\mathbf{A}_1 & \triangleq \begin{bmatrix} 0 & -v_x \\ 0 & 0 \end{bmatrix}, &&
\mathbf{A}_2  \triangleq \begin{bmatrix} -\dfrac{1}{m} & -\dfrac{1}{m} \\ -\dfrac{l_1}{I_z}& \dfrac{ l_2}{I_z} \end{bmatrix}, \nonumber\\
\mathbf{A}_3 & \triangleq \begin{bmatrix} \dfrac{2a_1}{\lambda_1} & \dfrac{2a_1l_1}{\lambda_1} \\  \dfrac{2a_2}{\lambda_2} & -\dfrac{2a_2l_2}{\lambda_2}\end{bmatrix}, &&
\mathbf{A}_4 \triangleq \begin{bmatrix} \dfrac{v_xC_1}{2a_1\lambda_1 w_1} & 0 \\ 0 & \dfrac{v_xC_2}{2a_2\lambda_2w_2} \end{bmatrix},
\end{align}
 and the input matrix $\mathbf{B} \in \mathbf{M}_2(\mathbb{R})$ is given by
\begin{align}\label{eq:B1Origin}
\mathbf{B} & \triangleq \begin{bmatrix} -\dfrac{2a_1v_x}{\lambda_1} & 0\\ 0 & -\dfrac{2a_2v_x}{\lambda_2}\end{bmatrix}.
\end{align}
The system \eqref{eq:ssOriginal}-\eqref{eq:B1Origin} is well-posed. In particular, consider the space $\mathcal{X} \triangleq \mathbb{R}^2 \times L^2((0,1);\mathbb{R}^2)$; then, \eqref{eq:ssOriginal} admits a unique \emph{mild solution} $(\bm{x},\bm{u}) \in C^0([0,T];\mathcal{X})$ for all ICs $(\bm{x}_0,\bm{u}_0) \in \mathcal{X}$ and inputs $\bm{\delta} \in L^2((0,T);\mathbb{R}^2)$ (see, e.g., \cite{Pazy,Tanabe}). 
The ODE-PDE interconnection \eqref{eq:ssOriginal}-\eqref{eq:B1Origin}, decomposed into its rigid body dynamics and tire dynamics subsystems, is schematized in Figure \ref{figure:system0}, where, for ease of visualization, $\mathbb{R}^2\ni \bm{U}_1(t) \triangleq \mathbf{A}_2\bm{Y}(t)$, and $\mathbb{R}^2 \ni \bm{U}_2(t) \triangleq \mathbf{A}_3\bm{x}(t) + \mathbf{B}\bm{\delta}(t)$.

The available measurements are instead assumed to be of the form $\mathbb{R}^{2} \ni \bm{y}(t) = [y_1(t) \; y_2(t)]^{\mathrm{T}}$, where $\mathbb{R} \ni y_1(t) = r(t)$, and
\begin{align}\label{eq:y2Poss}
y_2(t)  = \dod{u_1(\xi,t)}{t}, && \text{or} && y_2(t) = \dpd{u_1(0,t)}{\xi}.
\end{align}
Since $v_x$ and $a_1$ are assumed to be known, the measurement vector may thus be composed as
\begin{align}\label{eq:yC}
\bm{y}(t) = \mathbf{C}_1\bm{x}(t) + \mathbf{C}_2\bm{u}(1,t) + \mathbf{C}_3\bm{\delta}(t), 
\end{align}
with $\mathbf{C}_1, \mathbf{C}_2, \mathbf{C}_3 \in \mathbf{M}_2(\mathbb{R})$ reading
\begin{align}
\mathbf{C}_1 & \triangleq \begin{bmatrix} 0 & 1 \\ \dfrac{2a_1}{\lambda_1} & \dfrac{2a_1l_1}{\lambda_1} \end{bmatrix},\nonumber  \\
 \mathbf{C}_2 & \triangleq \begin{bmatrix} 0 & 0 \\ \dfrac{v_xC_1}{2a_1\lambda_1 w_1} & 0\end{bmatrix}, \nonumber \\
\mathbf{C}_3 & \triangleq \begin{bmatrix} 0 & 0 \\ -\dfrac{2a_1v_x}{\lambda_1} & 0\end{bmatrix}.
\end{align}

It is worth clarifying that the two possible expressions for $y_2(t)$, as reported in \eqref{eq:y2Poss}, correspond respectively to measuring the velocity of the bristle deformation traveling inside the contact patch of the front tires (at any arbitrary position $\xi \in [0,1]$), and the slope of the tire deformation at the leading edge. In particular, measurements of the first type may be obtained using accelerometers installed inside the tire, whereas measurements of the second type may be acquired again using accelerometers (by double integration of the acceleration signal), strain sensors, or optical devices \cite{Tuononen1,Tuononen2}. It should be stressed that, in this paper, only the front tires are supposed to be equipped with smart sensors (the design of the state observer carried out in Section \ref{sect:observerMain} further simplifies if the rear tires also equip similar devices). In fact, it is generally preferable to install intelligent tire technologies on the front axle rather than the rear, as variations in road friction conditions can be detected more rapidly. In the same context, it should be observed that, given the assumption of small slip angles in this study, measurements of the type \eqref{eq:y2Poss} could theoretically be obtained by instrumenting only one of the front tires. However, this approach may lack robustness in scenarios where operating conditions vary between the front tires, such as during a classic \text{\textmu}-split maneuver. Finally, a conclusive observation concerning the adoption of smart tire sensors is collected in Remark \ref{remark:1}.
\begin{remark}\label{remark:1}
Compared to measuring the slope of the tire deformation at the leading edge, using velocity measurements appears to be more appetible, since, according to the governing equations of the linear brush model \eqref{eq:PDEsfullcontST}, such a kinematic quantity is independent of the position of the bristle inside the contact patch. This allows employing a reduced number of sensors, which typically rotate with the tire itself and thus enter and relinquish the contact patch periodically. In practice, relying on velocity measurements requires that a sensor enters the contact patch before the preceding one leaves it. Thus, the number of needed sensors would realistically depend on the ratio between the tire radius and the contact patch length.
\end{remark}

Starting with \eqref{eq:ssOriginal}-\eqref{eq:yC}, the main objective of the present paper is formalized below. 
\begin{problem}[Tracking problem]\label{prob:track}
Consider the ODE-PDE system \eqref{eq:ssOriginal}-\eqref{eq:B1Origin}, along with a reference signal $\bm{x}\ped{ref} \in C^2([0,\infty];\mathbb{R}^{2}) \cap L^\infty([0,\infty);\mathbb{R}^2)$. Using the available measurements $\bm{y}(t) \in \mathbb{R}^{2}$, the \emph{tracking problem} consists of designing the control input $\mathbb{R}^2 \ni \bm{\delta}(t) = [\delta_1(t)\; \delta_2(t)]^{\mathrm{T}}$ so that $\bm{x}(t) \to \bm{x}\ped{ref}(t)$ for all ICs $(\bm{x}_0,\bm{u}_0)\in \mathcal{X}$, whilst ideally maintaining the PDE states bounded for all times.
\end{problem}
Exploiting the peculiar structure of the ODE-PDE interconnection \eqref{eq:ssOriginal}-\eqref{eq:B1Origin}, the above Problem \ref{prob:track} may be inferred to be equivalent to a typical force-control problem. In turn, the latter may be interpreted either as a simple stabilization problem, or alternatively as a path-following one (see Section \ref{sect:pathfoll}).

In this context, it is worth mentioning that, for a linear single-track model, the stabilization problem is not frequently encountered in classic vehicle dynamics, since vehicles are traditionally designed to be \emph{understeer} ($C_1 l_1 < C_2 l_2$), and hence stable for small values of the sideslip angle. However, the adoption of distributed models to describe tire dynamics has recently revealed the existence of dangerous Hopf bifurcations even for understeer vehicles, which motivates addressing Problem \ref{prob:track} using the formulation \eqref{eq:ssOriginal}--\eqref{eq:B1Origin} also in the case of a constant tracking signal $\bm{x}\ped{ref}(t) = \bm{x}^\star$ corresponding to an equilibrium. Specifically, the tracking Problem \ref{prob:track} is addressed in this manuscript by relying on some important stability properties of the ODE and PDE subsystems in isolation. To this end, the next Sections \ref{sect:equi} and \ref{sect:stability} recollect some preliminary results concerning the equilibria and stability of the interconnection \eqref{eq:ssOriginal}.

\subsection{System's equilibria}\label{sect:equi}
The stabilization of \eqref{eq:ssOriginal} requires calculating the equilibria $\bm{x}^\star, \bm{u}^\star(\xi) \in \mathbb{R}^2$ associated with a certain constant input $\bm{\delta} = \bm{\delta}^\star = [\delta_1^\star\; \delta_2^\star]^{\mathrm{T}}$.
Starting with the brush model equations \eqref{eq:PDEsfullcontST}, in stationary conditions, the distributed deformation of the tire inside the contact patch may be recovered as
\begin{align}\label{eq:brushSolSS}
u_i^\star(\xi) = 4\alpha_i\Bigl(v_y^\star, r^\star, \delta_i^\star\Bigr)\xi, && \xi \in [0,1],
\end{align}
for $i \in \{1,2\}$. Integrating \eqref{eq:brushSolSS} along the contact patch according to \eqref{eq:tireForce} provides the following expression for the steady-state axle forces
\begin{align}\label{eq:YSS}
Y_i^\star = C_i\alpha_i\Bigl(v_y^\star, r^\star, \delta_i^\star\Bigr), && i \in \{1,2\}.
\end{align}
In turn, substituting \eqref{eq:YSS} into \eqref{eq:bycicleVer1}, recalling \eqref{eq:lateralSlips}, and solving for $(v_y^\star,r^\star)$ yields
\begin{subequations}\label{eq:EquilibriaVyR}
\begin{align}
\begin{split}
v_y^\star & = \dfrac{v_xC_1C_2(l_1+l_2)(\delta_1^\star l_2+\delta_2^\star l_1)}{C_1C_2(l_1+l_2)^2-mv_x^2(C_1l_1-C_2l_2)} \\
& \quad -\dfrac{mv_x^3(C_1l_1\delta_1^\star-C_2l_2\delta_2^\star)}{C_1C_2(l_1+l_2)^2-mv_x^2(C_1l_1-C_2l_2)},
\end{split} \\
r^\star & = \dfrac{v_xC_1C_2(l_1+l_2)(\delta_1^\star-\delta_2^\star)}{C_1C_2(l_1+l_2)^2-mv_x^2(C_1l_1-C_2l_2)}.
\end{align}
\end{subequations}
As for the classic single-track model, the denominators in \eqref{eq:EquilibriaVyR} are always well-defined if the \emph{understeer} ($C_1l_1 < C_2l_2$) or \emph{neutral conditions} ($C_1l_1 = C_2l_2$) are satisfied. For, \emph{oversteer vehicles} ($C_1l_1 > C_2l_2$), \eqref{eq:EquilibriaVyR} predicts the existence of a critical longitudinal velocity
\begin{align}
v_x\ap{cr} & \triangleq \sqrt{\dfrac{C_1C_2(l_1+l_2)^2}{m(C_1l_1-C_2l_2)}}.
\end{align}
For such a value of longitudinal speed, no equilibrium exists; for $v_x \not = v_x\ap{cr}$, the equilibrium is clearly unique. The stability properties of the model \eqref{eq:ssOriginal} are investigated in Section \ref{sect:stability}.

\subsection{Stability}\label{sect:stability}
In the following, stability properties are enounced concerning both the whole ODE-PDE interconnection \eqref{eq:ssOriginal}-\eqref{eq:B1Origin} and the PDE subsystem \eqref{eq:ssOriginalPDE}-\eqref{eq:ssOriginalBC} in isolation. 

\subsubsection{Model stability}
To investigate the stability of the linear single-track model with distributed tires, it is beneficial to define the following matrices:
\begin{subequations}
\begin{align}\label{eq:SigmaMatr}
\begin{split}
\mathbf{\Sigma}(\xi,s) & \triangleq \int_{0}^{\xi}\exp\Bigl(-s\Upsilon^{-1}\bigl(\xi-\xi^\prime\bigr)\Bigr) \Upsilon^{-1}\dif \xi^\prime \\
& =\dfrac{1}{s} \begin{bmatrix} 1-\eu^{-\varsigma_1 s\xi } & 0 \\ 0 & 1-\eu^{-\varsigma_2 s\xi}\end{bmatrix}, \quad \xi \in [0,1],
\end{split} \\
\begin{split}
\mathbf{\Psi}(s) & \triangleq \int_0^1 \mathbf{\Sigma}(\xi,s)\dif \xi \\
& = \dfrac{1}{s^2}\begin{bmatrix} \dfrac{\varsigma_1 s + \eu^{-\varsigma_1 s}-1}{\varsigma_1} & 0 \\ 0 & \dfrac{\varsigma_2 s + \eu^{-\varsigma_2 s}-1}{\varsigma_2}\end{bmatrix},\label{eq:PsiMatr}
\end{split}
\end{align} 
\end{subequations}
with $\varsigma_i \triangleq 1/\upsilon_i = 2a_i/v_x$, $i \in \{1,2\}$. It is crucial to observe that $\mathbf{\Sigma}(\xi,\cdot), \mathbf{\Psi} \in H^\infty(\mathbb{C}_{>0};\mathbf{M}_2(\mathbb{C}))$ for any combination of model parameters.
The next result, formalized according to Lemma \ref{prop:stabilityChar}, provides a simple characterization for the stability of the ODE-PDE interconnection \eqref{eq:ssOriginal}-\eqref{eq:B1Origin}.

\begin{figure}
\centering
\includegraphics[width=1\linewidth]{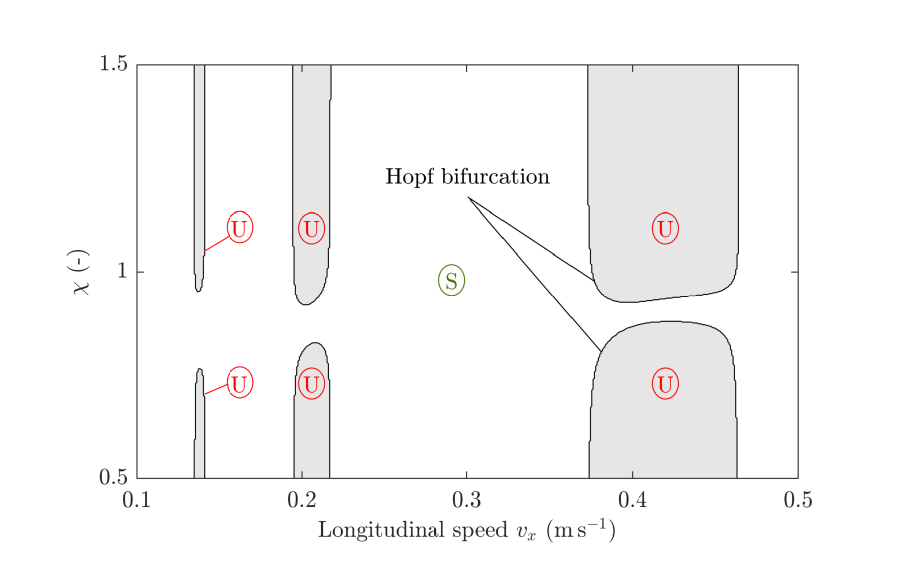} 
\caption{Stability charts for the single-track model with distributed tires, as described by \eqref{eq:ssOriginal}, for different values of the ratio $\chi \triangleq C_1l_1/(C_2l_2)$ and longitudinal speed $v_x$. The unstable regions (shaded) correspond to combinations of parameters for which $\det(A(s))$ has two roots with positive real part. Model parameter values as in Table \ref{table:Param1}.}
\label{figure:Bifurcation}
\end{figure}

\begin{figure}
\centering
\includegraphics[width=1\linewidth]{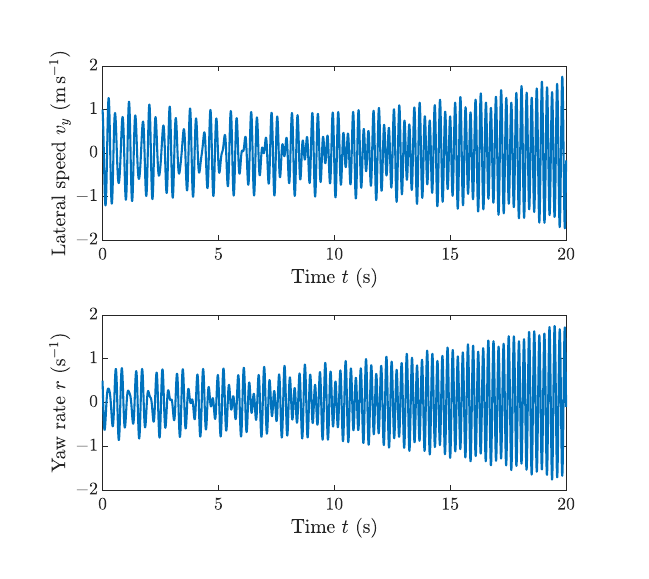} 
\caption{Typical shimmy behavior of an understeer vehicle traveling at low speed ($v_x = 0.4$ $\text{m}\,\text{s}^{-1}$). Model parameter values as in Table \ref{table:Param1}.}
\label{figure:Bifurcation2}
\end{figure}

\begin{lemma}[Stability of the linear single-track model with distributed tires]\label{prop:stabilityChar}
Consider the ODE-PDE interconnection described by \eqref{eq:ssOriginal}-\eqref{eq:B1Origin}, along with the matrix 
\begin{align}\label{eq:A(p)}
\mathbf{A}(s) \triangleq \begin{bmatrix}s\mathbf{I}_2-\mathbf{A}_1 & -\mathbf{A}_2 & \bm{0}\\
-\mathbf{\Psi}(s)\mathbf{A}_3 & \mathbf{I}_2 & -\mathbf{\Psi}(s)\mathbf{A}_4 \\
-\mathbf{\Sigma}(1,s)\mathbf{A}_3 & \bm{0} &  \mathbf{I}_2-\mathbf{\Sigma}(1,s)\mathbf{A}_4\end{bmatrix}, 
\end{align}
where $\mathbf{\Sigma}(1,s)$ and $\mathbf{\Psi}(s)$ are defined according to \eqref{eq:SigmaMatr} and \eqref{eq:PsiMatr}, resepctively.
Then, if $\det\bigl(\mathbf{A}(s)\bigr) \not = 0$ for all $s \in \mathbb{C}_{\geq 0}$, the system \eqref{eq:ssOriginal}-\eqref{eq:B1Origin} is stable. 
\begin{proof}
See \cite{BicyclePDE}.
\end{proof}
\end{lemma}
Specifically, the conditions stated in Lemma \ref{prop:stabilityChar} ensure stability in the norm $\norm{\cdot}_{\mathcal{X}}$. From the numerical experiments conducted in \cite{BicyclePDE}, it may be conjectured that, for sufficiently large longitudinal speeds, the requirement enounced in Lemma \ref{prop:stabilityChar} be equivalent to the classic understeer condition $C_1 l_1 < C_2 l_2$ formulated for the standard version of the single-track model. On the other hand, as opposed to the classic linear single-track model, the version with distributed tires predicts the existence of oscillatory instabilities occurring at low longitudinal velocities (typically $v_x \leq 5$ $\text{m}\,\text{s}^{-1}$), possibly related to Hopf bifurcations \cite{BicyclePDE}. These unstable behaviors, associated with micro-shimmy oscillations documented also experimentally, induce increased energy consumption and slip losses; addressing Problem \ref{prob:track} becomes thus crucial. Figure \ref{figure:Bifurcation} illustrates the stability chart of a single-track model with distributed tires driving at low speed, for the nondimensional parameter $\chi \triangleq C_1l_1/(C_2l_2)$ varying between 0.5 and 1.5. The shaded areas represent the unstable regions associated with micro-shimmy phenomena. As opposed to what might intuitively be expected, an inspection of Figure \ref{figure:Bifurcation} confirms the existence of unstable oscillations also for very understeer vehicles, whereas mildly oversteer behaviors seem to be unaffected by such a type of instability. 
Figure \ref{figure:Bifurcation2} also plots the trend of the lateral speed and yaw rate for an understeer vehicle driving at $v_x = 0.4$ $\text{m}\,\text{s}^{-1}$. The combination of model parameters and cruising speed corresponds to the rightmost shaded region in Figure \ref{figure:Bifurcation}, associated with the micro-shimmy phenomenon. As it can be observed from Figure \ref{figure:Bifurcation2}, both kinematic quantities start spontaneously oscillating, with amplitude increasing over time.
Figures \ref{figure:Bifurcation} and \ref{figure:Bifurcation2} were produced numerically using standard parameter values from the literature, reported in Table \ref{table:Param1}.

Unfortunately, the condition on $\det(\mathbf{A}(s))$ in Lemma \ref{prop:stabilityChar} needs to be checked numerically for a given set of parameter values. Conversely, the stability of the PDE subsystem \eqref{eq:ssOriginalPDE}-\eqref{eq:ssOriginalBC} may be proved rigorously for any combination of model parameters, as done in \cite{BicyclePDE,MioStand}, as well as in Section \ref{sect:PDEstab}. 

\subsubsection{Stability of the PDE subsystem}\label{sect:PDEstab}
The exponential stability of the PDEs \eqref{eq:ssOriginalPDE} follows directly from Lemma \ref{lemma:stabPDE} below.
\begin{lemma}\label{lemma:stabPDE}
Consider the following scalar PDE:
\begin{subequations}\label{eq:PDEscalar}
\begin{align}
\begin{split}
&\dpd{u(\xi,t)}{t} + \upsilon\dpd{u(\xi,t)}{t} = \gamma u(1,t) + \beta(t), \\
& \qquad \qquad \qquad \qquad \qquad \quad(\xi,t) \in (0,1)\times (0,T), \end{split} \label{eq:PDEgeneral} \\
&u(0,t) = 0, \quad t \in (0,T),
\end{align}
\end{subequations}
with parameters $(\upsilon,\gamma) \in \mathbb{R}_{>0}\times\mathbb{R}$ satisfying $\abs{\bar{\gamma}} \triangleq \dfrac{\abs{\gamma}}{\upsilon} \in (0,1]$. Then, the transfer function $G_{\widehat{\beta}(s) \to \widehat{u}(\xi,s)}(s)$ from $\widehat{\beta}(s)$ to $\widehat{u}(\xi,s)$, given by
\begin{align}\label{eq:Gtrasf}
G_{\widehat{\beta}(s) \to \widehat{u}(\xi,s)}(s) & = \dfrac{\varsigma \zeta(\xi,s)}{1-\bar{\gamma}\zeta(1,s)},
\end{align}
with $\mathbb{R}_{>0}\ni \varsigma \triangleq 1/\upsilon$ and
\begin{align}
\zeta(\xi,s) & \triangleq \dfrac{1-\eu^{-\varsigma s\xi}}{\varsigma s},
\end{align}
is stable for all $\abs{\bar{\gamma}} \in (0,1)$, i.e., $G_{\widehat{\beta}(s) \to \widehat{u}(\xi,s)} \in H^\infty(\mathbb{C}_{>0};\mathbb{C})$; moreover, it has a single unstable pole in zero for $\bar{\gamma} = 1$.
\begin{proof}
Taking the Laplace transform of \eqref{eq:PDEscalar} and solving for $\widehat{u}(\xi,s)$ yields
\begin{align}\label{eq:Lapl1}
\widehat{u}(\xi,s) = \zeta(\xi,s)\Bigl[ \bar{\gamma}\widehat{u}(1,s) + \varsigma\widehat{\beta}(s)\Bigr].
\end{align}
In turn, computing $\widehat{u}(1,s)$ provides
\begin{align}\label{eq:Lapl2}
\widehat{u}(1,s) & = \dfrac{\varsigma \zeta(1,s)}{1-\bar{\gamma}\zeta(1,s)}\widehat{\beta}(s).
\end{align}
Combining \eqref{eq:Lapl1} with \eqref{eq:Lapl2} finally gives \eqref{eq:Gtrasf}. Since $\norm{\zeta(\xi,\cdot)}_\infty \leq 1$ for all $\xi \in [0,1]$, the transfer function $G_{\widehat{\beta}(s) \to \widehat{u}(\xi,s)}(s)$ is stable for all $\bar{\gamma}$ satisfying $\abs{\bar{\gamma}} \in (0,1)$. Furthermore, defining $\rho \triangleq \bar{\gamma}\eu^{-\bar{\gamma}}$, the poles of $G_{\widehat{\beta}(s) \to \widehat{u}(\xi,s)}(s)$ may be determined as
\begin{align}\label{eq:solS}
s = \upsilon\Bigl[ W_{k}(\rho\eu^{\im \text{\textpi}})+\bar{\gamma}\Bigr], \quad k \in \mathbb{Z}.
\end{align}
where $W_{k}(\cdot)$, $k \in \mathbb{Z}$, denotes the branch $k$ of the Lambert function $W : \mathbb{C} \to \mathbb{C}$. For $\bar{\gamma} = 1$, \eqref{eq:solS} immediately reveals the existence of a single pole in zero.
\end{proof}
\end{lemma}
The two PDEs \eqref{eq:ssOriginalPDE} are coupled together only via the lumped states. Therefore, by recalling the definition of relaxation lengths $\lambda_i$, $i \in \{1,2\}$, using \eqref{eq:relaxLength}, it is easy to conclude that both satisfy the assumptions of Lemma \ref{lemma:stabPDE} with $\bar{\gamma} \in (0,1)$, and are thus exponentially stable when considered in isolation. This observation is crucial for the synthesis of the observer carried out in Section \ref{sect:observer}. 

\section{State-feedback control}\label{sect:StateFeed}
The tracking of a reference trajectory for the kinematic variables may be conveniently achieved via force control. 
To this end, decomposing the lumped and distributed states as $\bm{x}(t) = \bm{x}\ped{ref}(t) + \bm{x}_\delta(t)$, $\bm{u}(\xi,t) = \bm{u}\ped{ref}(\xi,t) + \bm{u}_\delta(\xi,t)$, the equations governing the dynamics of the tracking error may be cast as\footnote{Here, the subscript $(\cdot)_\delta$, not to be confounded with the notation adopted for the steering wheel input, indicates the error variables.}
\begin{subequations}\label{eq:perturbed}
\begin{align}
\begin{split}
& \dot{\bm{x}}_\delta(t) = \mathbf{A}_1\bigl(\bm{x}_\delta(t)+\bm{x}\ped{ref}(t)\bigr)  \\
& \qquad \quad +\mathbf{A}_2\bigl(\bm{Y}_\delta(t) + \bm{Y}\ped{ref}(t)\bigr) -\dot{\bm{x}}\ped{ref}(t), \quad t \in (0,T), \label{eq:OdesubsDelta}
\end{split}\\
\begin{split}
& \dpd{\bm{u}_\delta(\xi,t)}{t} + \mathbf{\Upsilon} \dpd{\bm{u}_\delta(\xi,t)}{\xi} = \mathbf{A}_3\bigl(\bm{x}_{\delta}(t)+\bm{x}\ped{ref}(t)\bigr) \\
& \qquad \quad+ \mathbf{A}_4\bigl(\bm{u}_\delta(1,t)+\bm{u}\ped{ref}(1,t)\bigr)  -  \dpd{\bm{u}\ped{ref}(\xi,t)}{t} \\
& \qquad \quad - \mathbf{\Upsilon} \dpd{\bm{u}\ped{ref}(\xi,t)}{\xi}+ \mathbf{B}\bm{\delta} (t), \quad (\xi,t) \in (0,1) \times (0,T), \label{eq:perturbedPDEcompl}
\end{split}\\
& \bm{u}_\delta(0,t) = \bm{0}, \quad t \in (0,T).
\end{align}
\end{subequations}
The next step consists of computing the dynamics of the tracking error axle forces as a function of the reference ones, by defining
\begin{subequations}
\begin{align}
\bm{Y}\ped{ref}(t) &\triangleq \begin{bmatrix}\dfrac{C_1}{2a_1} & 0 \\ 0 &  \dfrac{C_2}{2a_2}\end{bmatrix}\int_0^1 \bm{u}\ped{ref}(\xi,t) \dif \xi, \label{eq:Y_deltau_deltaRelTrack}\\
\bm{Y}_\delta(t) & \triangleq \begin{bmatrix}\dfrac{C_1}{2a_1} & 0 \\ 0 &  \dfrac{C_2}{2a_2}\end{bmatrix}\int_0^1 \bm{u}_\delta(\xi,t) \dif \xi, && t \in [0,T),\label{eq:Y_deltau_deltaRel}
\end{align}
\end{subequations}
so that $\bm{Y}(t) = \bm{Y}\ped{ref}(t) + \bm{Y}_\delta (t)$.
Consequently, differentiating \eqref{eq:Y_deltau_deltaRel} with respect to the time, substituting the PDE dynamics \eqref{eq:perturbedPDEcompl}, and using the definition \eqref{eq:Y_deltau_deltaRelTrack} yields
\begin{align}\label{eq:Yref}
\begin{split}
\dot{\bm{Y}}_\delta(t) & = \bar{\mathbf{A}}_3\bigl(\bm{x}_\delta(t)+\bm{x}\ped{ref}(t)\bigr) + \bar{\mathbf{A}}_4\bigl(\bm{u}_\delta(1,t)+\bm{u}\ped{ref}(1,t)\bigr) \\
& \quad + \bar{\mathbf{B}}\bm{\delta}(t) -\dot{\bm{Y}}\ped{ref}(t), \quad t\in (0,T),
\end{split}
\end{align}
with  $\bar{\mathbf{A}}_3, \bar{\mathbf{A}}_4 \in \mathbf{M}_2(\mathbb{R})$ reading
\begin{align}\label{eq:Abar}
\bar{\mathbf{A}}_3 & \triangleq \begin{bmatrix} \dfrac{C_1}{\lambda_1} & \dfrac{C_1 l_1}{\lambda_1} \\ \dfrac{C_2}{\lambda_2} & -\dfrac{C_2 l_2}{\lambda_2}\end{bmatrix}, && \text{and} && \bar{\mathbf{A}}_4  \triangleq \begin{bmatrix} -\dfrac{v_xC_1}{4a_1\lambda_1} & 0 \\ 0 & -\dfrac{v_xC_2}{4a_2\lambda_2}\end{bmatrix},
\end{align}
and $\bar{\mathbf{B}} \in \mathbf{GL}_2(\mathbb{R})$ given by
\begin{align}
\bar{\mathbf{B}} & \triangleq \begin{bmatrix} -\dfrac{v_xC_1}{\lambda_1}& 0 \\ 0 &-\dfrac{v_xC_2}{\lambda_2} \end{bmatrix}.
\end{align}
Starting with \eqref{eq:Yref}, the signal $\dot{\bm{Y}}\ped{ref}(t) \in \mathbb{R}^2$ needs to be opportunely specified to ensure the asymptotic convergence of the state $\bm{x}_\delta(t)$ to zero. Proposition \ref{prop:stabODE} asserts the stabilizability of both the ODE subsystems \eqref{eq:ssOriginalODE} and \eqref{eq:Yref}.

\begin{proposition}[Stabilizability of the ODE systems \eqref{eq:ssOriginalODE} and \eqref{eq:Yref}]\label{prop:stabODE}
The pairs $(\mathbf{A}_1,\mathbf{A}_2)$ and $(\mathbf{0},\bar{\mathbf{B}})$ are stabilizable.
\begin{proof}
The result immediately follows by noting that $\det(\mathbf{A}_2) = -(l_1+l_2)/(I_zm)$ and $\det(\bar{\mathbf{B}}) = v_x^2C_1C_2/(\lambda_1\lambda_2)$.
\end{proof}
\end{proposition} 
In the presence of full-state measurements, Proposition \ref{prop:stabODE} permits designing the control input to suppress the coupling terms between the ODE and PDE equations, whilst simultaneously ensuring the stabilization of the lumped system \eqref{eq:ssOriginalODE}. Indeed, Proposition \ref{prop:stabODE} additionally implies the invertibility of the matrix $\bar{\mathbf{B}} \in \mathbf{GL}_2(\mathbb{R})$, which is required to decouple the dynamics of the variable $\bm{Y}_\delta(t)$ from those of $\bm{x}(t)$ and $\bm{u}(1,t)$.
Using \eqref{eq:OdesubsDelta}, \eqref{eq:Yref} and \eqref{eq:Abar}, a state-feedback controller may be designed as described in Lemma \ref{lemma:contrYrefState}.
\begin{lemma}\label{lemma:contrYrefState}
Consider the ODE-PDE system \eqref{eq:ssOriginal}, along with the tracking error dynamics \eqref{eq:OdesubsDelta} and \eqref{eq:Yref}. Then, the control input
\begin{align}\label{eq:controlInputY}
\begin{split}
\bm{\delta}(t) & = \bar{\mathbf{F}}\bigl(\bm{Y}(t)-\bm{Y}\ped{ref}(t)\bigr) \\
& \quad -\bar{\mathbf{B}}^{-1}\Bigl[\bar{\mathbf{A}}_3\bm{x}(t)+\bar{\mathbf{A}}_4\bm{u}(1,t)-\dot{\bm{Y}}\ped{ref}(t)\Bigr],
\end{split}
\end{align}
with
\begin{subequations}\label{eq:compatibilityDIsStar}
\begin{align}
\bm{Y}\ped{ref}(t) & \triangleq \mathbf{F}\bigl(\bm{x}(t)-\bm{x}\ped{ref}(t)\bigr)-\mathbf{A}_2^{-1}\bigl(\mathbf{A}_1\bm{x}\ped{ref}(t)-\dot{\bm{x}}\ped{ref}(t)\bigr),\label{eq:Ystar} \\
\begin{split}
\dot{\bm{Y}}\ped{ref}(t) & \triangleq \mathbf{F}\bigl(\mathbf{A}_1\bm{x}(t) + \mathbf{A}_2\bm{Y}(t)-\dot{\bm{x}}\ped{ref}(t)\bigr)\\
& \quad -\mathbf{A}_2^{-1}\bigl(\mathbf{A}_1\dot{\bm{x}}\ped{ref}(t)-\ddot{\bm{x}}\ped{ref}(t)\bigr), \label{eq:YstarDot}
\end{split}
\end{align}
\end{subequations}
and $\mathbf{F},\bar{\mathbf{F}}\in \mathbf{M}_2(\mathbb{R})$ chosen such that $\mathbf{M}_2(\mathbb{R})\ni \mathbf{A}_1^\prime \triangleq \mathbf{A}_1 + \mathbf{A}_2\mathbf{F}$ and $ \bar{\mathbf{B}}\bar{\mathbf{F}} \in \mathbf{M}_2(\mathbb{R})$ are Hurwitz, ensures that $\norm{(\bm{x}_\delta(t), \bm{Y}_\delta(t))}_2 \to 0$ exponentially fast for all ICs $(\bm{x}_0,\bm{u}_0) \in \mathcal{X}$.
\begin{proof}
Substituting \eqref{eq:controlInputY} into \eqref{eq:Yref} provides
\begin{align}
\dot{\bm{Y}}_\delta(t) & = \bar{\mathbf{B}}\bar{\mathbf{F}}\bm{Y}_\delta(t), \quad t\in(0,T),
\end{align}
which is exponentially stable by assumption. Therefore, $\norm{\bm{Y}_\delta(t)}_2\to 0$ exponentially fast. Furthermore, inserting \eqref{eq:Ystar} into \eqref{eq:OdesubsDelta} yields
\begin{align}\label{eqeq}
\dot{\bm{x}}_\delta(t) = \mathbf{A}_1^\prime \bm{x}_\delta(t) + \mathbf{A}_2\bm{Y}_\delta(t), \quad t \in (0,T).
\end{align}
Since $\mathbf{A}_1^\prime$ is Hurwitz by assumption, and $\bm{Y}_\delta(t) \to \bm{0}$ exponentially fast, $\norm{\bm{x}_\delta(t)}_2 \to 0$ exponentially fast. Combining the two assertions provides the desired result.
\end{proof}
\end{lemma}
Lemma \ref{lemma:contrYrefState} above asserts the exponential convergence of the ODE states and axle forces $(\bm{x}(t),\bm{Y}(t))$ to the desired reference signals $(\bm{x}\ped{ref}(t),\bm{Y}\ped{ref}(t))$. However, no statement is made concerning the distributed states. In practice, the compatibility condition \eqref{eq:compatibilityDIsStar} ensures that the distributed states converge asymptotically to zero. In the same context, it is perhaps worth observing that the term $\mathbf{B}_2\bar{\mathbf{B}}^{-1}\dot{\bm{Y}}\ped{ref}(t) \in \mathbb{R}^2$ may be interpreted as a reference velocity signal. The result is formalized in Lemma \ref{corollary2} below.
\begin{lemma}\label{corollary2}
Under the same assumptions of Lemma \ref{lemma:contrYrefState}, suppose additionally that $\dot{\bm{Y}}\ped{ref} \in L^1((0,\infty);\mathbb{R}^2)$; then, $\norm{\bm{u}(\cdot,t)}_{L^2((0,1);\mathbb{R}^2)}$ remains bounded for all times. Moreover, consider the decomposition $\bm{u}(\xi,t) \triangleq \bm{u}\ped{ref}(\xi,t) + \bm{u}_\delta(\xi,t)$, with $\bm{u}\ped{ref}(\xi,t) \in \mathbb{R}^2$ satisfying
\begin{subequations}\label{eq:uStar}
\begin{align}
\begin{split}
& \dpd{\bm{u}\ped{ref}(\xi,t)}{t} + \mathbf{\Upsilon} \dpd{\bm{u}\ped{ref}(\xi,t)}{\xi} = \mathbf{\Upsilon} \bm{u}\ped{ref}(1,t) - \mathbf{B}\bar{\mathbf{B}}^{-1}\dot{\bm{Y}}\ped{ref}(t),\\
& \qquad \qquad \qquad \qquad \qquad \qquad \quad (\xi,t) \in (0,1)\times(0,T), 
\end{split} \\
& \bm{u}\ped{ref} (0,t) = \bm{0}, \quad t \in (0,T).
\end{align}
\end{subequations}
Then, $\bm{u}_\delta(\xi,t) \to \bm{0}$ asymptotically.
\begin{proof}
Desiging the control input according to \eqref{eq:controlInputY}, the PDE subsystem becomes
\begin{subequations}\label{eq:PDEwithInput}
\begin{align}
\begin{split}
& \dpd{\bm{u}(\xi,t)}{t} + \mathbf{\Upsilon} \dpd{\bm{u}(\xi,t)}{\xi} = \mathbf{\Upsilon} \bm{u}(1,t)  + \mathbf{B}\bar{\mathbf{F}}\bm{Y}_\delta(t) \\
& \qquad \qquad \qquad \qquad \qquad \quad -\mathbf{B}\bar{\mathbf{B}}^{-1}\dot{\bm{Y}}\ped{ref}(t), \\
& \qquad \qquad \qquad \qquad \qquad \qquad \quad (\xi,t) \in (0,1)\times(0,T), 
\end{split}\\
& \bm{u} (0,t) = \bm{0}, \quad t \in (0,T).
\end{align}
\end{subequations} 
According to Lemma \ref{lemma:stabPDE}, the above PDE \eqref{eq:PDEwithInput} is marginally stable, since its transfer function has a single unstable pole in zero, and the remaining ones with negative real part. Thus, $\bm{Y}_\delta, \dot{\bm{Y}}\ped{ref} \in L^1((0,\infty);\mathbb{R}^2)$ implies that the PDE states remain bounded for all times. Consider now the decomposition $\bm{u}(\xi,t) \triangleq \bm{u}\ped{ref}(\xi,t) + \bm{u}_\delta(\xi,t)$, with $\bm{u}\ped{ref}(\xi,t) \in \mathbb{R}^2$ solving \eqref{eq:uStar}; the dynamics of the PDE error states is then governed by
\begin{subequations}
\begin{align}
\begin{split}
& \dpd{\bm{u}_\delta(\xi,t)}{t} + \mathbf{\Upsilon} \dpd{\bm{u}_\delta(\xi,t)}{\xi} = \mathbf{\Upsilon} \bm{u}_\delta(1,t)  + \mathbf{B}\bar{\mathbf{F}}\bm{Y}_\delta(t), \\
& \qquad \qquad \qquad \qquad \qquad \qquad \quad (\xi,t) \in (0,1)\times(0,T), 
\end{split}\\
& \bm{u}_\delta (0,t) = \bm{0}, \quad t \in (0,T).
\end{align}
\end{subequations} 
This is again a marginally stable PDE, with its transfer function containing a single unstable pole in zero. Since $\bm{Y}_\delta(t) \to \bm{0}$ exponentially fast, $\bm{u}_\delta(\xi,t)$ must converge to a steady-state function function $\mathbb{R}^2 \ni \bar{\bm{u}}_\delta(\xi) \triangleq \lim_{t \to \infty} \bm{u}_\delta(\xi,t)$ satisfying
\begin{subequations}
\begin{align}
& \dod{\bar{\bm{u}}_\delta(\xi)}{\xi} = \bar{\bm{u}}_\delta(1), \quad \xi \in (0,1), \\
& \bar{\bm{u}}_\delta(0) = \bm{0}.
\end{align}
\end{subequations}
The unique solution to the above ODE reads obviously
\begin{align}
\bar{\bm{u}}_\delta(\xi) = \bar{\bm{u}}_\delta(1)\xi, \quad \xi \in [0,1].
\end{align}
But \eqref{eq:Y_deltau_deltaRel} immediately implies that
\begin{align}
 \int_0^1 \bar{\bm{u}}_\delta(\xi) \dif \xi = \lim_{t \to \infty} \begin{bmatrix} \dfrac{2a_1}{C_1} & 0 \\ 0 & \dfrac{2a_2}{C_2}\end{bmatrix} \bm{Y}_\delta(t) = \bm{0},
\end{align}
which gives $\bar{\bm{u}}_\delta(1) = \bm{0}$ and hence $\bar{\bm{u}}_\delta(\xi) = \bm{0}$ for all $\xi \in [0,1]$. This concludes the proof.
\end{proof}
\end{lemma}
The interpretation of Lemma \ref{corollary2} is as follows: the distributed error state $\bm{u}_\delta(\xi,t)$ decays asymptotically to zero, leaving the tracked signal $\mathbf{B}_2\bar{\mathbf{B}}^{-1}\dot{\bm{Y}}\ped{ref}(t)$ as the sole contributor to tire deformation. From this analysis, it follows that if $\bm{Y}\ped{ref}(t)$ converges exponentially to a steady-state value, the bristle deformation will also converge asymptotically. This scenario occurs, for example, when addressing a pure stabilization problem around a known equilibrium, where the bristle deformations within the contact patch converge to the steady-state value corresponding to the stationary axle forces. It should be noted that the condition $\dot{\bm{Y}}\ped{ref} \in L^1((0,\infty);\mathbb{R}^2)$ may be reformulated in terms of $\bm{x}\ped{ref}(t)$ and its derivatives, and can therefore be ensured by an opportune specification of desired trajectories. 

Lemmata \ref{lemma:contrYrefState} and \ref{corollary2} offer a complete solution to the tracking Problem \ref{prob:track} when assuming full-state measurements. In practice, the lateral speed and the slip angles are not simultaneously measured and must be estimated. The observer design constitutes the scope of Section \ref{sect:observerMain}.

\section{Observer design and output-feedback controller}\label{sect:observerMain}
The present Section is devoted to the synthesis of a state estimator, along with the design of an output-feedback controller.

\subsection{Observer design}\label{sect:observer}
The method outlined in the following relies on decoupling the PDE subsystem \eqref{eq:ssOriginalPDE} from the ODE one. Specifically, the proposed technique exploits the exponential stability of the PDEs \eqref{eq:ssOriginalPDE}, which ensures the convergence of the estimated distributed states to the true values; then, the ODE subsystem \eqref{eq:ssOriginalODE} is stabilized in isolation. In this context, the detectability of the lumped equation is first asserted in Proposition \ref{prop:detect}.
\begin{proposition}[Detectability of the ODE subsystem \eqref{eq:ssOriginalODE}]\label{prop:detect}
The pair $(\mathbf{A}_1,\mathbf{C}_1)$ is detectable.
\begin{proof}
The result immediately follows by noting that $\det(\mathbf{C}_1) = -2a_1/\lambda_1$.
\end{proof}
\end{proposition}
In fact, Propositon \ref{prop:detect} also implies the invertibility of the matrix $\mathbf{C}_1 \in \mathbf{GL}_2(\mathbb{R})$, which is essential in decoupling the PDE and ODE subsystems.
With this premises, denoting respectively with $\hat{\bm{x}}(t) \in \mathbb{R}^2$, $\hat{\bm{u}}(\xi,t) \in \mathbb{R}^2$, and $\hat{\bm{y}}(t) \in \mathbb{R}^{2}$ the estimates of $\bm{x}(t)$, $\bm{u}(\xi,t)$, and $\bm{y}(t)$, the following observer is proposed:
\begin{subequations}\label{eq:ObserverXXX}
\begin{align}
\begin{split}
& \dot{\hat{\bm{x}}}(t) = \mathbf{A}_1 \hat{\bm{x}}(t) +\mathbf{A}_2\hat{\bm{Y}}(t) \\
& \qquad \quad  - \mathbf{L}_1\bigl(\bm{y}(t)-\hat{\bm{y}}(t)\bigr), \quad t \in (0,T), 
\end{split}\\
\begin{split}
& \dpd{\hat{\bm{u}}(\xi,t)}{t} + \mathbf{\Upsilon} \dpd{\hat{\bm{u}}(\xi,t)}{\xi} = \mathbf{A}_3 \hat{\bm{x}}(t) + \mathbf{A}_4 \hat{\bm{u}}(1,t)  + \mathbf{B}\bm{\delta}(t)\\
& \qquad \qquad \qquad \qquad \qquad \quad -\mathbf{L}_2\bigl(\bm{y}(t)-\hat{\bm{y}}(t)\bigr),  \\
& \qquad \qquad \qquad \qquad \qquad \quad (\xi,t) \in (0,1) \times (0,T), \\
& \hat{\bm{u}}(0,t) = \bm{0}, \quad t \in (0,T),
\end{split}
\end{align}
\end{subequations}
where
\begin{align}
\hat{\bm{Y}}(t) \triangleq \begin{bmatrix}\dfrac{C_1}{2a_1} & 0 \\ 0 & \dfrac{C_2}{2a_2} \end{bmatrix}\int_0^1 \hat{\bm{u}}(\xi,t) \dif \xi, && t\in[0,T),
\end{align}
and
\begin{align}
\hat{\bm{y}}(t) = \mathbf{C}_1\hat{\bm{x}}(t) + \mathbf{C}_2\hat{\bm{u}}(1,t) + \mathbf{C}_3\bm{\delta}(t).
\end{align}
Defining the errors $\mathbb{R}^2 \ni \tilde{\bm{x}}(t) \triangleq \bm{x}(t)-\hat{\bm{x}}(t)$, $\mathbb{R}^2 \ni \tilde{\bm{u}}(\xi,t) \triangleq \bm{u}(\xi,t) - \hat{\bm{u}}(\xi,t)$, and $\mathbb{R}^{2} \ni \tilde{\bm{y}}(t) = \bm{y}(t)-\hat{\bm{y}}(t)$, the observer error dynamics may be deduced to obey
\begin{subequations}\label{eq:Observer}
\begin{align}
\begin{split}
& \dot{\tilde{\bm{x}}}(t) = \mathbf{A}_1 \tilde{\bm{x}}(t) +\mathbf{A}_2\tilde{\bm{Y}}(t) + \mathbf{L}_1\tilde{\bm{y}}(t), \; t \in (0,T), 
\end{split} \label{eq:obsODE}\\
\begin{split}
& \dpd{\tilde{\bm{u}}(\xi,t)}{t} + \mathbf{\Upsilon} \dpd{\tilde{\bm{u}}(\xi,t)}{\xi} = \mathbf{A}_3 \tilde{\bm{x}}(t) + \mathbf{A}_4 \tilde{\bm{u}}(1,t) +\mathbf{L}_2\tilde{\bm{y}}(t),\\
& \qquad \qquad \qquad \qquad \qquad \quad (\xi,t) \in (0,1) \times (0,T), 
\end{split} \label{eq:obsPDE}\\
& \tilde{\bm{u}}(0,t) = \bm{0}, \quad t \in (0,T),
\end{align}
\end{subequations}
clearly with
\begin{align}
\begin{split}
\tilde{\bm{Y}}(t) & \triangleq \bm{Y}(t)-\hat{\bm{Y}}(t) \\
& = \begin{bmatrix}\dfrac{C_1}{2a_1} & 0 \\ 0 & \dfrac{C_2}{2a_2} \end{bmatrix}\int_0^1 \tilde{\bm{u}}(\xi,t) \dif \xi, \quad t\in[0,T).
\end{split}
\end{align}
Concerning instead the output error, the following equation may be derived:
\begin{align}\label{eq:yTilde}
\tilde{\bm{y}}(t) = \mathbf{C}_1\tilde{\bm{x}}(t) + \mathbf{C}_2\tilde{\bm{u}}(1,t).
\end{align}
Finally, due to the invertibility of the matrix $\mathbf{C}_1$, \eqref{eq:yTilde} also leads to
\begin{align}\label{eq:Xinv}
\tilde{\bm{x}}(t) = \mathbf{C}_1^{-1}\bigl[\tilde{\bm{y}}(t)-\mathbf{C}_2\tilde{\bm{u}}(1,t)\bigr],
\end{align}
which allows expressing $\tilde{\bm{x}}(t)$ as a function of the available measurement error, plus a linear combination of distributed state errors. As anticipated above, \eqref{eq:Xinv} may be conveniently exploited to decouple the PDE subsystem \eqref{eq:obsPDE} from the ODE one \eqref{eq:obsODE}. The result is formalized according to Theorem \ref{th:obs} below.

\begin{theorem}\label{th:obs}
Consider the observer \eqref{eq:ObserverXXX}, along with the error dynamics \eqref{eq:Observer}, and assume that $\mathbf{L}_1 \in \mathbf{M}_2(\mathbb{R})$ is chosen such that $\mathbf{M}_2(\mathbb{R})\ni \mathbf{A}_1^* \triangleq \mathbf{A}_1 + \mathbf{L}_1\mathbf{C}_1$ is Hurwitz, and $\mathbf{L}_2 \in \mathbf{M}_2(\mathbb{R})$ is selected as $\mathbf{L}_2 = -\mathbf{A}_3\mathbf{C}_1^{-1}$.
Then, $\norm{(\tilde{\bm{x}}(t), \tilde{\bm{u}}(\cdot,t))}_{\mathcal{X}}, \norm{\tilde{\bm{Y}}(t)}_2 \to 0$ exponentially fast for all ICs $(\bm{x}_0,\bm{u}_0), (\hat{\bm{x}}_0,\hat{\bm{u}}_0) \in \mathcal{X}$.
\begin{proof}
The gain $\mathbf{L}_2 = -\mathbf{A}_3\mathbf{C}_1^{-1}$ yields the following cascaded PDE system:
\begin{subequations}\label{eq:transp0}
\begin{align}
& \dpd{\tilde{u}_1(\xi,t)}{t} + \dfrac{v_x}{2a_1}\dpd{\tilde{u}_1(\xi,t)}{\xi} = 0, \label{eq:transp1}\\
\begin{split}
& \dpd{\tilde{u}_2(\xi,t)}{t} + \dfrac{v_x}{2a_2}\dpd{\tilde{u}_2(\xi,t)}{\xi} = \dfrac{v_xC_2}{2a_2\lambda_2w_2}\tilde{u}_2(1,t)\\
& \qquad \qquad \qquad \qquad \qquad \qquad  -\dfrac{a_2C_1v_x}{2a_1^2\lambda_2w_1}\tilde{u}_1(1,t), \\
& \qquad \qquad \qquad \qquad \qquad \qquad \quad (\xi,t)\in(0,1)\times(0,T),
\end{split}\label{eq:transp2} \\
& \tilde{\bm{u}}(0,t) = \bm{0}, \quad t \in (0,T).\label{eq:BCutilde}
\end{align}
\end{subequations}
In particular, the first PDE \eqref{eq:transp1} consists of a simple transport equation whose state converges to zero after a finite time $\bar{t}_1 \triangleq 2a_1/v_x$. The second PDE \eqref{eq:transp2} is exponentially stable due to Lemma \ref{lemma:stabPDE}. Therefore, since $\tilde{u}_1(1,t) = 0$ for all $t \geq \bar{t}_1$, it may be concluded that $\norm{\tilde{u}_2(\xi,t)}_{L^2((0,1);\mathbb{R})} \to 0$ exponentially fast for all $\tilde{\bm{u}}_0 \in H^1((0,1);\mathbb{R}^2)$ satisfying the BC \eqref{eq:BCutilde}, which also implies $\norm{\tilde{\bm{Y}}(t)}_2 \to 0$ exponentially fast. Finally, the ODE subsystem rewrites
\begin{align}\label{eq:ODExassb}
\tilde{\bm{x}}(t) = \mathbf{A}_1^*\tilde{\bm{x}}(t) + \mathbf{A}_2\tilde{\bm{Y}}(t) + \mathbf{L}_1\mathbf{C}_2\tilde{\bm{u}}(1,t), && t\in(0,T).
\end{align}
Consequently, since $\mathbf{A}_1^*$ is Hurwitz by assumption, and $\tilde{\bm{Y}}(t), \tilde{\bm{u}}(1,t)\to 0$ exponentially fast, $\norm{\tilde{\bm{x}}(t)}_2 \to 0$ exponentially fast. Combining the previous assertions yields the result. The generalization to ICs $\tilde{\bm{u}}_0 \in L^2((0,1);\mathbb{R}^{2})$ follows from standard density arguments.
\end{proof}
\end{theorem}

In conjunction with the control strategy developed in Section \ref{sect:StateFeed}, the observer \eqref{eq:ObserverXXX} may be used to synthesize an output-feedback controller addressing the tracking Problem \ref{prob:track}. 

\subsection{Output-feedback backstepping controller design}
In the following the estimates of $\bm{Y}\ped{ref}(t)$ and $\dot{\bm{Y}}\ped{ref}(t)$ are indicated respectively with $\hat{\bm{Y}}\ped{ref}(t) \in \mathbb{R}^2$ and $\hat{\dot{\bm{Y}}}\ped{ref}(t) \in \mathbb{R}^2$; the corresponding errors are denoted by $\mathbb{R}^2 \ni \tilde{\bm{Y}}\ped{ref}(t) = \bm{Y}\ped{ref}(t)-\hat{\bm{Y}}\ped{ref}(t)$, and $\mathbb{R}^2\ni \tilde{\dot{\bm{Y}}}\ped{ref}(t) = \dot{\bm{Y}}\ped{ref}(t)-\hat{\dot{\bm{Y}}}\ped{ref}(t)$. 

The main result of the paper is delivered by Theorem \ref{th:outputFeed} below.
\begin{theorem}\label{th:outputFeed}
Consider the ODE-PDE system \eqref{eq:ssOriginal}, along with the tracking error dynamics \eqref{eq:OdesubsDelta} and \eqref{eq:Yref}, and the observer \eqref{eq:ObserverXXX}, under the same assumptions of Lemma \ref{lemma:contrYrefState} and Theorem \ref{th:obs}. Then, the control input
\begin{align}\label{eq:controlInputYFeed}
\begin{split}
\bm{\delta}(t) & = \bar{\mathbf{F}}\bigl(\hat{\bm{Y}}(t)-\hat{\bm{Y}}\ped{ref}(t)\bigr) \\
& \quad -\bar{\mathbf{B}}^{-1}\Bigl[\bar{\mathbf{A}}_3\hat{\bm{x}}(t)+\bar{\mathbf{A}}_4\hat{\bm{u}}(1,t)-\hat{\dot{\bm{Y}}}\ped{ref}(t)\Bigr],
\end{split}
\end{align}
with
\begin{subequations}\label{eq:compatibilityDIsStarFeed}
\begin{align}
\hat{\bm{Y}}\ped{ref}(t) & \triangleq \mathbf{F}\bigl(\hat{\bm{x}}(t)-\bm{x}\ped{ref}(t)\bigr)-\mathbf{A}_2^{-1}\bigl(\mathbf{A}_1\bm{x}\ped{ref}(t)-\dot{\bm{x}}\ped{ref}(t)\bigr), \\
\begin{split}
\hat{\dot{\bm{Y}}}\ped{ref}(t) & \triangleq \mathbf{F}\bigl(\mathbf{A}_1\hat{\bm{x}}(t) + \mathbf{A}_2\hat{\bm{Y}}(t)-\dot{\bm{x}}\ped{ref}(t)\bigr)\\
& \quad -\mathbf{A}_2^{-1}\bigl(\mathbf{A}_1\dot{\bm{x}}\ped{ref}(t)-\ddot{\bm{x}}\ped{ref}(t)\bigr), \label{eq:YstarDotFeed}
\end{split}
\end{align}
\end{subequations}
ensures that $\norm{\bm{x}_\delta(t)}_2, \norm{\bm{Y}_\delta(t)}_2 \to 0$ exponentially fast for all ICs $(\bm{x}_0,\bm{u}_0), (\hat{\bm{x}}_0,\hat{\bm{u}}_0) \in \mathcal{X}$.
\begin{proof}
The observer error dynamics is still given by \eqref{eq:Observer}; hence, Theorem \ref{th:obs} implies that $\norm{(\tilde{\bm{x}}(t),\tilde{\bm{u}}(\cdot,t))}_{\mathcal{X}} \to 0$ exponentially fast for all $(\tilde{\bm{x}}_0,\tilde{\bm{u}}_0) \in \mathbb{R}^2 \times H^1((0,1);\mathbb{R}^2)$ satisfying the BC \eqref{eq:BCutilde}. Moreover, with the proposed control law, the ODE \eqref{eq:Yref} becomes
\begin{align}\label{TOutputFeedback}
\begin{split}
\dot{\bm{Y}}_\delta(t) & = \bar{\mathbf{B}}\bar{\mathbf{F}}\bm{Y}_\delta(t)-\bar{\mathbf{B}}\bar{\mathbf{F}}\bigl(\tilde{\bm{Y}}(t)-\tilde{\bm{Y}}\ped{ref}(t)\bigr) + \mathbf{A}_3\tilde{\bm{x}}(t) \\
& \quad + \mathbf{A}_4\tilde{\bm{u}}(1,t) -\tilde{\dot{\bm{Y}}}\ped{ref}(t), \quad t\in(0,T),
\end{split}
\end{align}
with
\begin{align}\label{eq:THHHH}
\tilde{\bm{Y}}\ped{ref}(t) & = \mathbf{F}\tilde{\bm{x}}(t), && \text{and} && \tilde{\dot{\bm{Y}}}\ped{ref}(t) = \mathbf{F}\bigl(\mathbf{A}_1\tilde{\bm{x}}(t) + \mathbf{A}_2\tilde{\bm{Y}}(t)\bigr).
\end{align}
Since $\bar{\mathbf{B}}\bar{\mathbf{F}}$ is Hurwitz by assumption, and $\tilde{\bm{Y}}_\delta(t), \tilde{\bm{x}}(t), \tilde{\bm{u}}(1,t), \tilde{\dot{\bm{Y}}}\ped{ref}(t) \to \bm{0}$ exponentially fast for all $(\tilde{\bm{x}}_0,\tilde{\bm{u}}_0) \in \mathbb{R}^2\times H^1((0,1);\mathbb{R}^2)$, $\norm{\bm{Y}_\delta(t)}_2\to 0$ exponentially fast. Furthermore, inserting \eqref{eq:Ystar} yields again \eqref{eqeq}, implying that $\norm{\bm{x}_\delta(t)}_2 \to 0$ exponentially fast. Combining the two assertions provides the desired result.The generalization to ICs $(\tilde{\bm{x}}_0,\tilde{\bm{u}}_0) \in \mathcal{X}$ follows from standard density arguments.
\end{proof}
\end{theorem}
Finally, concerning the distributed states, Lemma \ref{corollary3} represents the equivalent of \ref{corollary2} for the output-feedback tracking problem.

\begin{lemma}\label{corollary3}
Under the same assumptions of Theorem \ref{th:outputFeed}, suppose additionally that $\dot{\bm{Y}}\ped{ref} \in L^1((0,\infty);\mathbb{R}^2)$; then, $\norm{\bm{u}(\cdot,t)}_{L^2((0,1);\mathbb{R}^2)}$ remains bounded for all times. Moreover, consider the same decomposition $\bm{u}(\xi,t) \triangleq \bm{u}\ped{ref}(\xi,t) + \bm{u}_\delta(\xi,t)$, with $\bm{u}\ped{ref}(\xi,t) \in \mathbb{R}^2$ as in Lemma \ref{corollary2}.
Then, $\bm{u}_\delta(\xi,t) \to \bm{0}$ asymptotically.
\begin{proof}
The result may be proved using a similar rationale as that adopted in the proof of Lemma \ref{corollary2}. 
\end{proof}
\end{lemma}

In conjunction with Lemma \ref{corollary3}, Theorem \ref{th:outputFeed} completely solves the tracking Problem \ref{prob:track} in the case of partial state measurements. The performance of the proposed observer and controller is exemplified in Section \ref{sect:Val}.


\section{Simulation and numerical validation}\label{sect:Val}
The controller and the observer developed in Section \ref{sect:observerMain} are finally tested considering two different scenarios: stabilization in the presence of micro-shimmy oscillations, and path tracking via force control.

The plant model utilized herein is a higher-fidelity double-track vehicle model that incorporates the effects of lateral load transfers (see, e.g., Equation (3.158) in \cite{Guiggiani}, Chapter 3), along with a nonlinear representation of the distributed tire dynamics, as opposed to the linear formulation introduced in Section \ref{sect:MDesc}. Notably, the refined model accurately captures the influence of finite Coulomb friction, resulting in saturated tire characteristics even under transient conditions. For a comprehensive discussion of this formulation, readers are referred, for instance, to \cite{LibroMio} (Chapter 4), or \cite{CarcassDyn}. The parameter values for the single-track model utilized for observed and controller design are instead listed in Table \ref{table:Param1}.
\begin{table}[]
\centering
\begin{tabular}{llll}
\toprule
\textbf{Parameter} & \textbf{Description}           & \textbf{Value} & \textbf{Unit}             \\ \midrule
$m$                & Vehicle mass                   & 1300           & kg                        \\
$I_z$              & Moment of inertia              & 2000           & $\text{kg}\,\text{m}^3$   \\
$l_1$              & Front axle length              & 1              & m                         \\
$l_2$              & Rear axle length               & 1.6            & m                         \\
$C_1$           & Front axle stiffness           & $7\cdot 10^4$  & $\text{N}$ \\
$C_2$           & Rear axle stiffness            & $9\cdot 10^4$  & $\text{N}$ \\
$a_1$              & Front patch semilength & 0.055          & m                         \\
$a_2$              & Rear patch semilength  & 0.045          & m                         \\
$\lambda_1$         & Front relaxation length        & 0.195          & m                         \\
$\lambda_2$         & Rear relaxation length         & 0.225          & m                 \\
\bottomrule       
\end{tabular}
\caption{Parameter values for the linear single-track vehicle model with distributed tires.}\label{table:Param1}
\end{table}


\subsection{Suppression of micro-shimmy oscillations}\label{sect:stabilizing}

\begin{figure*}
\centering
\includegraphics[width=0.9\linewidth]{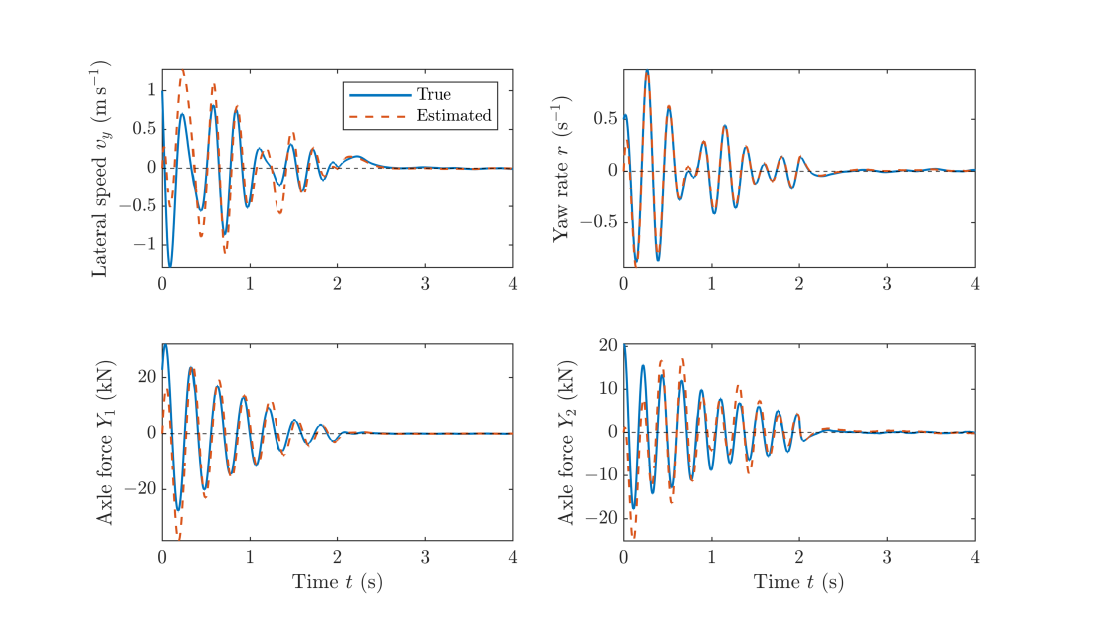} 
\caption{True (solid blue line) and estimated (dashed orange line) states for the nonlinear double-track model operating in an unstable parameter region associated with micro-shimmy oscillations, and subjected to the control input \eqref{eq:controlInputYFeed} with $\bm{x}\ped{ref}(t) = \bm{x}^\star = \bm{0}$, after $t = 2$ s.}
\label{figure:BifurcationSuppressed}
\end{figure*}

As anticipated in Sections \ref{sect:Intro} and \ref{sect:stability}, micro-shimmy oscillations may arise at low cruising speeds in both understeer and oversteer vehicles. This phenomenon excites unpredictable dynamic behaviors that can compromise handling stability, and cause increased energy losses. Hence, the first scenario analyzed in the paper concerns the suppression of micro-shimmy oscillations in an understeer vehicle traveling at a constant longitudinal speed $v_x = 0.4$ $\text{m}\text{s}^{-1}$. In the absence of steering input, and with the same model parameters as those used to produce Figure \ref{figure:Bifurcation}, the vehicle starts spontaneously oscillating. Figure \ref{figure:BifurcationSuppressed} plots the trend of the true and estimated states obtained when the input is designed according to \eqref{eq:controlInputYFeed}, with $\bm{x}\ped{ref}(t) = \bm{x}^\star = \bm{0}$.
Specifically, the controller intervenes only after $t = 2$ s, a sufficient time for detecting the presence of micro-shimmy oscillations. For this test scenario, as for the subsequent ones, the yaw rate and smart tire sensor measurements, as well as the steering wheel inputs, are added with white noise with typical characteristics of those found in standard automotive instrumentation, whereas the observer and controller gains are selected as
\begin{align}
\mathbf{L}_1 & = \begin{bmatrix}45 & -44.3 \\ -40 & 0 \end{bmatrix}, \nonumber \\ 
\mathbf{F} & = 10^3 \cdot \begin{bmatrix} 24 & 7.08 \\ 15 & -33.1 \end{bmatrix}, \nonumber \\
 \bar{\mathbf{F}} & = 10^{-6}\cdot\begin{bmatrix}2.8 & 0 \\ 0 & 2.5 \end{bmatrix}.
\end{align}
In spite of the presence of additive white noise, the designed control action successfully stabilizes the system in a relatively short time, around $t = 4$ s.

\begin{figure*}
\centering
\includegraphics[width=0.9\linewidth]{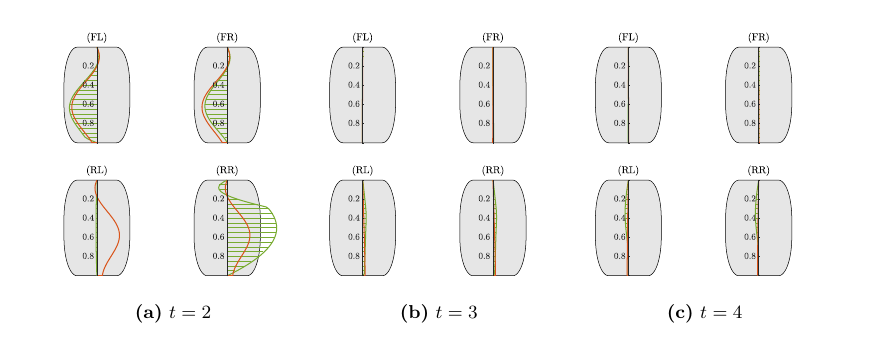} 
\caption{Actual (green) and estimated (orange) distributed tire deformation inside the contact patch for the micro-shimmy stabilizing maneuver described in Section \ref{sect:stabilizing}: \textbf{(a)} $t = 2$ s; \textbf{(b)} $t = 3$ s; \textbf{(c)} $t = 4$ s.}
\label{tiresCombined1}
\end{figure*}

Figure \ref{tiresCombined1} illustrates the trend of the actual (green) and estimated\footnote{The distributed variables appearing in the PDEs \eqref{eq:PDEsfullcontST} represent the \emph{equivalent} bristle deformation of each axle, that is, the sum of the deflections of the left and right tires. Owing to the assumption of linearity, the deformation on the left and right tires of the same axle must however be the same. Therefore, the estimated bristle deformation for each of the two tires mounted on the same axle is obtained by dividing the distributed variables in \eqref{eq:PDEsfullcontST} by two.} (orange) distributed bristle deformation inside the tires' contact patch, for three different values of time $t = 2$, 3, and 4 s. In particular, the large deformations depicted in Figure \ref{tiresCombined1}\textbf{(a)} refer to the instant immediately preceding the intervention of the controller, when the vehicle still exhibits dangerous micro-shimmy behavior. In this context, it is interesting to observe how the front tires, albeit undergoing relatively large deflections, both operate in the linear region; on the other hand, the rear tires are dynamically saturated. This is especially evident concerning the rear right tire, where the direction of the bristles' deformation gradually changes, and sliding already occurs around $\xi = 0.3$. The rear left tire is also saturated, but the effect is less evident due to the substantial decrease in normal force and available friction produced by the load transfer. In theoretical accordance with Lemma \ref{corollary3}, the controller also suppresses the transient deflection of the bristles, as it may be inferred directly by inspection of Figures \ref{tiresCombined1}\textbf{(b)} and \ref{tiresCombined1}\textbf{(c)}, where the front tires are completely unloaded, whereas the rear ones are subjected to small and oscillating deformations. These residual deformations are to be ascribed to the persistent noise added to the measurements and steering wheel input, as confirmed by simulation results.


\subsection{Path-following via force control}\label{sect:pathfoll}
Path-following and force allocation problems are typically addressed offline using model predictive control, with the resulting solution subsequently employed as the reference signal for the controller during online operation \cite{Fors1,Fors2,Fors3,Fors4}. In contrast, this paper proposes the generation of the reference signal in real-time, concurrently with the control action. Owing to this premise, the reference value $\bm{x}\ped{ref}(t)$ for the kinematic variables may be deduced starting with the equations describing the vehicle's trajectory:
\begin{subequations}\label{eq:trajectory}
\begin{align}
\dot{x}_O(t) & = v_x\cos\bigl(\psi(t)\bigr)-v_y(t)\sin\bigl(\psi(t)\bigr), \label{eq:trajectory11}\\ 
\dot{y}_O(t) & = v_x\sin\bigl(\psi(t)\bigr) + v_y(t)\cos\bigl(\psi(t)\bigr), \label{eq:trajectory12}\\
\dot{\psi}(t) & = r(t), && t \in (0,T).
\end{align}
\end{subequations}
Omitting the speed assignment problem, the above \eqref{eq:trajectory11} and \eqref{eq:trajectory12} are subsequently linearized around zero yaw angles and lateral velocities, providing
\begin{subequations}\label{eq:trajectory2}
\begin{align}
\dot{x}_O(t) & \approx v_x, \\ 
\dot{y}_O(t) & \approx v_x\psi(t) + v_y(t), && t \in (0,T).
\end{align}
\end{subequations}
In contrast to \eqref{eq:trajectory11} and \eqref{eq:trajectory12}, \eqref{eq:trajectory2} do not permit explicit control of the vehicle's longitudinal position, but allow to deduce a simple law for the reference values of $y_O(t)$ and $\psi(t)$, denoted respectively as $y_{O,\textnormal{ref}}(t), \psi\ped{ref}(t) \in \mathbb{R}$. In fact, according to \eqref{eq:trajectory2}, is sufficient to specify
\begin{subequations}
\begin{align}
v_{y,\textnormal{ref}}(t) & = -f_1\bigl(y_O(t)-y_{O,\textnormal{ref}}(t)\bigr)-v_x\psi\ped{ref}(t) + \dot{y}_{O,\textnormal{ref}}(t), \\
r\ped{ref}(t) & =-f_2\bigl(\psi(t)-\psi\ped{ref}(t)\bigr) + \dot{\psi}\ped{ref}(t), 
\end{align}
\end{subequations}
for some positive constants $f_1, f_2 \in \mathbb{R}_{>0}$, and adequately smooth signals $(y_{O,\textnormal{ref}}, \psi\ped{ref}) \in C^3([0,\infty);\mathbb{R}^2) \cap L^\infty([0,\infty);\mathbb{R}^2)$. In particular, these are postulated as\footnote{Notably, owing to the aassumption of small steering angles, the actual longitudinal position $x_O(t)$ in \eqref{eq:sineMan} may conveniently be approximated as $x_O(t) \approx v_x t$, as also done in \eqref{eq:trajectory2}.} 
\begin{subequations}\label{eq:sineMan}
\begin{align}
y_{O,\textnormal{ref}}(t) & = A_y\sin\bigl(\omega_y x_O(t)\bigr), \\
\psi\ped{ref}(t) & = A_\psi\cos\bigl(\omega_\psi x_O(t)\bigr),
\end{align}
\end{subequations}
with $A_y,A_\psi,\omega_y,\omega_\psi \in \mathbb{R}_{>0}$.
Choosing, for instance, $\omega_\psi = \omega_y \triangleq \omega$ and $A_\psi = \omega A_y$, the vehicle attempts to follow a sinusoidal path whilst approximately maintaining tangency to it. Furthermore, the gains $f_1$ and $f_2$ are selected as
\begin{align}
f_1 & = 8, && \text{and} && f_2 = 3,
\end{align}
whereas two different sets of values for the parameters $(A_y, \omega)$ are prescribed to emulate different types of maneuvers.

\begin{figure*}
\centering
\includegraphics[width=0.9\linewidth]{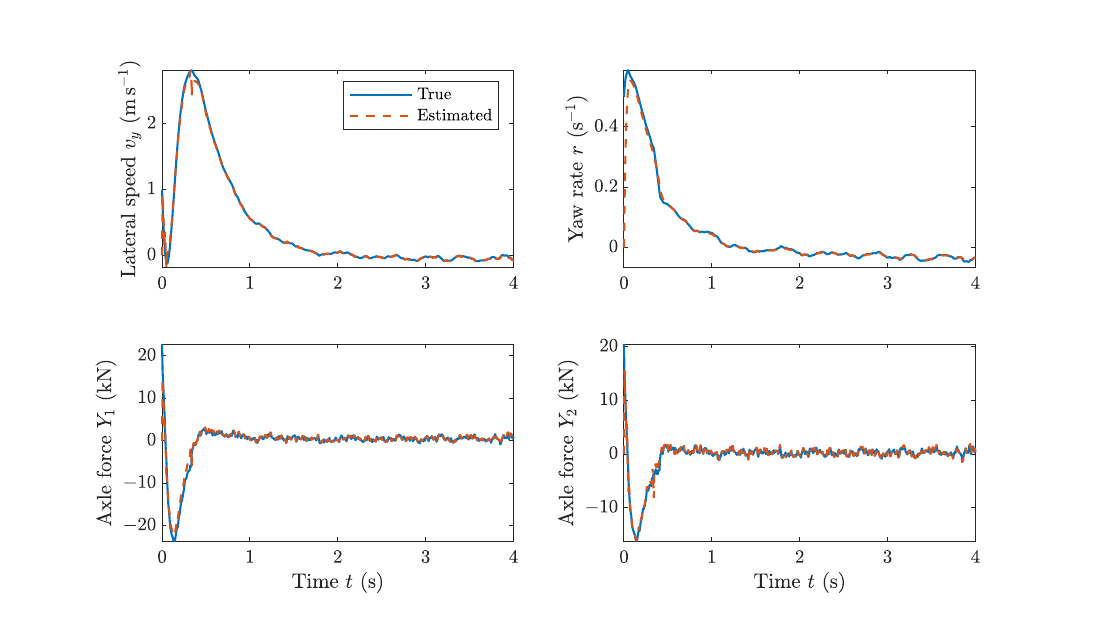} 
\caption{True (solid blue line) and estimated (dashed orange line) states for the nonlinear double-track model performing the sine maneuver \eqref{eq:sineMan} with $A_y = 25$ m, and $\omega = 0.01$ $\text{m}^{-1}$, and subjected to the control input \eqref{eq:controlInputYFeed}.}
\label{fobserverTraj}
\end{figure*}

\begin{figure}
\centering
\includegraphics[width=1\linewidth]{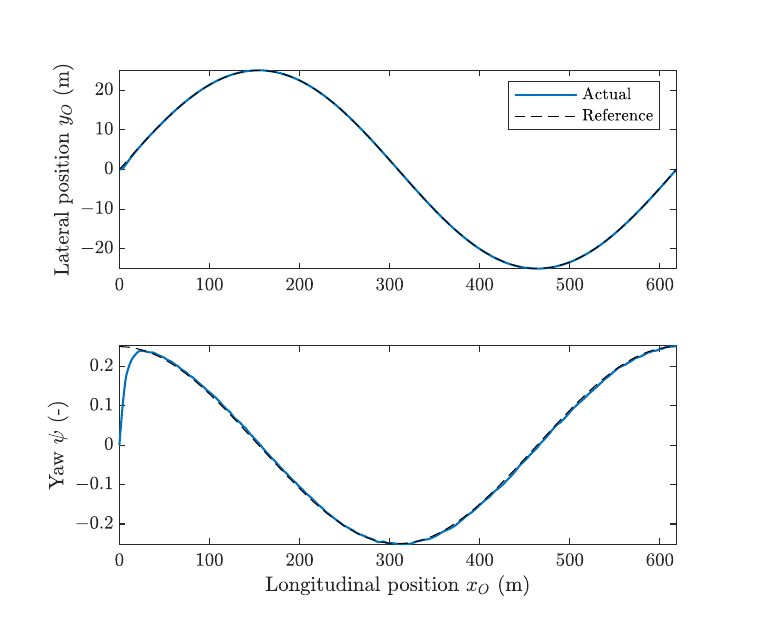} 
\caption{Actual and reference trajectories for the vehicle's lateral position and yaw for the sine maneuver \eqref{eq:sineMan} with $A_y = 25$ m, and $\omega = 0.01$ $\text{m}^{-1}$.}
\label{Traj}
\end{figure}

More specifically, the first case analyzed in the following imitates a mild sine maneuver where the vehicle drives at a constant longitudinal speed of $v_x = 20$ $\text{m}\,\text{s}^{-1}$, and negotiates two consecutive symmetric curves over a relatively long distance, amounting approximately to 60 m. Such a scenario corresponds to $A_y = 25$ m, and $\omega = 0.01$ $\text{m}^{-1}$. 
The comparison between the actual and estimated kinematic variables, and axle forces, is illustrated in Figure \ref{fobserverTraj} for the first four seconds of simulation. In all the reported cases, the observer converges rapidly to the true values, and tracks them satisfactorily as they evolve over time, with a small noise affecting only the estimates of the axle forces. The actual and reference trajectories for the vehicle's lateral position $y_O(t)$ and yaw $\psi(t)$ are instead plotted in Figure \ref{Traj}. In particular, the output-feedback controller manages to track the prescribed trajectory with great accuracy. 

\begin{figure*}
\centering
\includegraphics[width=0.9\linewidth]{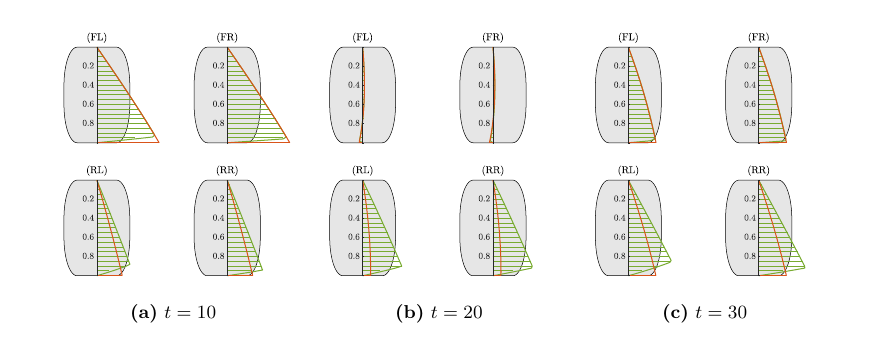} 
\caption{Actual (green) and estimated (orange) distributed tire deformation inside the contact patch for the sine maneuver \eqref{eq:sineMan} with $A_y = 25$ m, and $\omega = 0.01$ $\text{m}^{-1}$: \textbf{(a)} $t = 10$ s; \textbf{(b)} $t = 20$ s; \textbf{(c)} $t = 30$ s.}
\label{tiresCombined2}
\end{figure*}

The deformation undergone by the four tires is depicted in Figure \ref{tiresCombined2} for $t = 10$, 20, and 30 s. In practice, with the designed control action, the axle forces oscillate around zero to track the sinusoidal trajectory, which explains the corresponding trend observed in Figure \ref{tiresCombined2}. In particular, for $t = 10$ s, the front tires appear to work almost entirely in the linear region, whereas the rear ones manifest the presence of small sliding regions in the proximity of the contact patches' trailing edges. Moreover, the estimated bristle deflection matches almost perfectly the true one in the front tires, whereas a little discrepancy may be noticed for the rear axle, where the assumption of linearity is obviously violated. The situation appears significantly different in Figure \ref{tiresCombined2}\textbf{(b)}, where the front tires are essentially unloaded, whereas the rear ones undergo major deformations, which are however not captured by the observer. Concerning instead Figure \ref{tiresCombined2}\textbf{(c)}, similar considerations hold as for \ref{tiresCombined2}\textbf{(a)}, but with a reversed load proportion between the front and rear axle. Also in this case, the observer partially fails to predict the nonlinear trend of the bristle deflection.

\begin{figure*}
\centering
\includegraphics[width=0.9\linewidth]{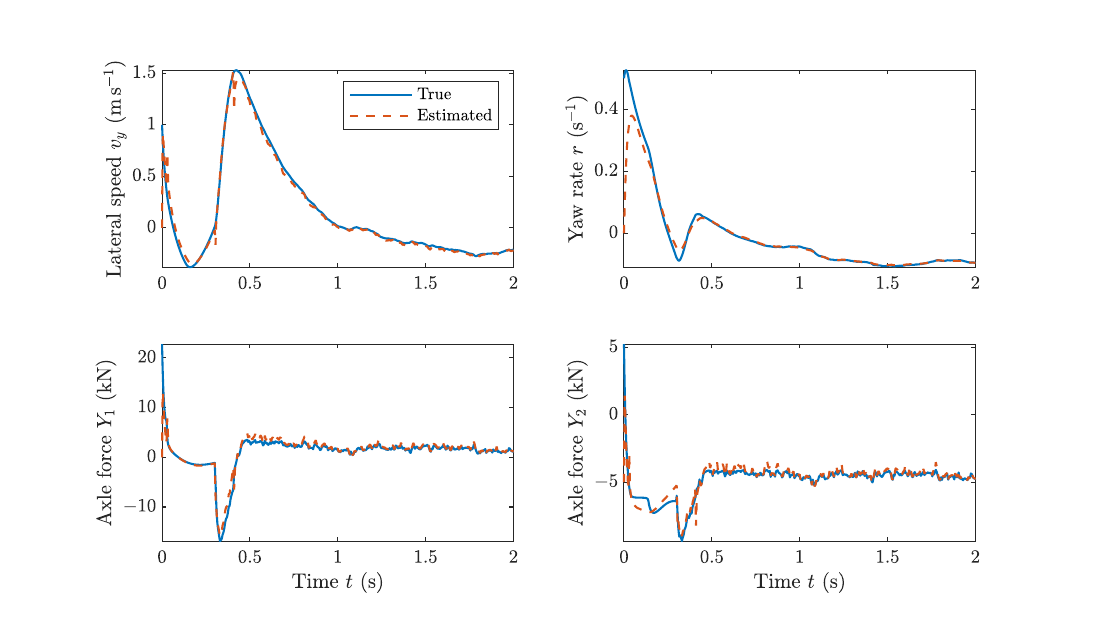} 
\caption{True (solid blue line) and estimated (dashed orange line) states for the nonlinear double-track model performing the sine maneuver \eqref{eq:sineMan} with $A_y = 2$ m, and $\omega = 0.05$ $\text{m}^{-1}$, and subjected to the control input \eqref{eq:controlInputYFeed} after $t = 0.3$ s.}
\label{fobserverTraj2}
\end{figure*}

\begin{figure}
\centering
\includegraphics[width=1\linewidth]{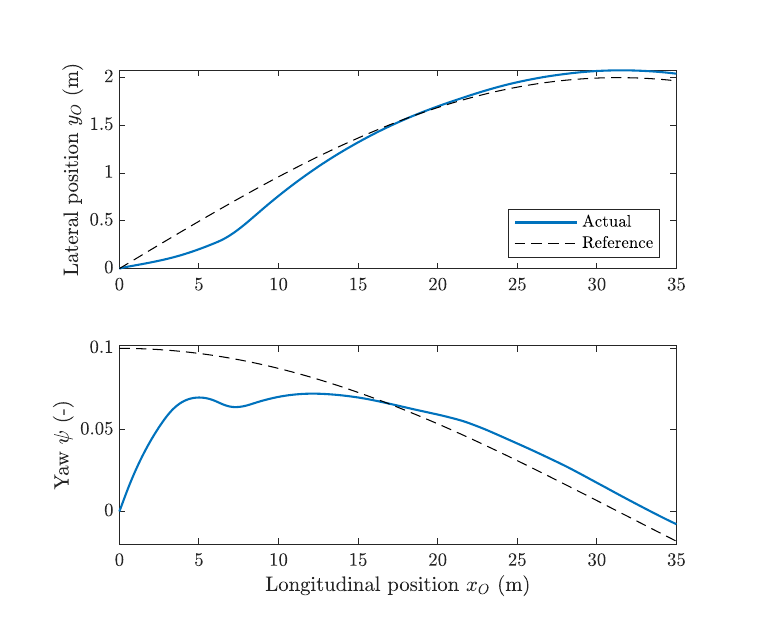} 
\caption{Actual and reference trajectories for the vehicle's lateral position and yaw for the sine maneuver \eqref{eq:sineMan} with $A_y = 2$ m, and $\omega = 0.05$ $\text{m}^{-1}$.}
\label{Traj2}
\end{figure}

The final scenario investigated in this Section refers to a more aggressive maneuver that resembles a typical collision avoidance action, similar to those described, for instance, in \cite{Fors1,Fors2,Fors3,Fors4}. Considering again a vehicle driving at $v_x = 20$ $\text{m}\,\text{s}^{-1}$, the parameters in \eqref{eq:sineMan} are thus specified as $A_y = 2$ m, and $\omega = 0.05$ $\text{m}^{-1}$. These values correspond to avoiding an obstacle positioned 30 m ahead of the vehicle, with an approximate width of 1.5 m. To provide a more realistic representation of the test scenario, the controller is activated only after $t=0.3$ s, which is a reasonable delay for an autonomous vehicle. Figures \ref{fobserverTraj2} and \ref{Traj2} report respectively the trend of the true and estimated states, and the trajectories (actual and reference) for the vehicle's lateral position and yaw. Notably, the rapid intervention of the output-feedback controller effectively prevents a collision, despite minor deviations from the reference yaw trajectory. These deviations are deemed acceptable, as the primary objective of the controller is to ensure collision avoidance, and the variations in yaw remain minimal. Specifically, the maximum deviations in the final part of the maneuver amount to 7.4 cm and 0.011 radians for the lateral position and yaw, respectively.

\begin{figure*}
\centering
\includegraphics[width=0.9\linewidth]{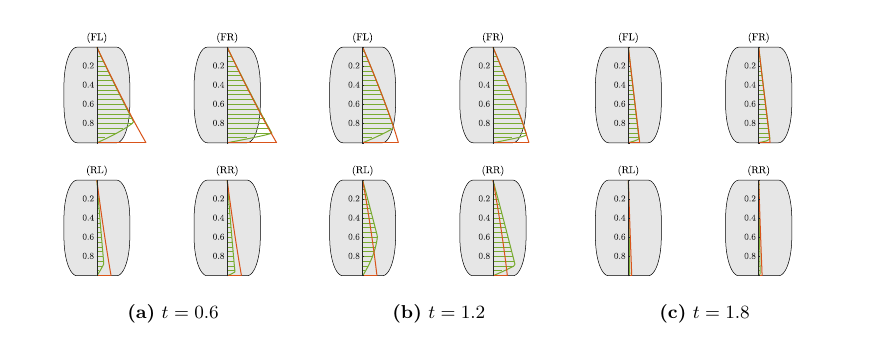} 
\caption{Actual (green) and estimated (orange) distributed tire deformation inside the contact patch for the sine maneuver \eqref{eq:sineMan} with $A_y = 2$ m, and $\omega = 0.05$ $\text{m}^{-1}$: \textbf{(a)} $t = 0.6$ s; \textbf{(b)} $t = 1.2$ s; \textbf{(c)} $t = 1.8$ s.}
\label{tiresCombined3}
\end{figure*}

The bristle deflection within the tires' contact patches is finally illustrated in Figure \ref{tiresCombined3} for $t = 0.6$, 1.2, and 1.8 s. Specifically, $t = 0.6$ corresponds to twice the intervention time of the controller, whereas $t = 1.8$ to that needed by the vehicle to overcome the obstacle. Generally speaking, the conclusions that may be drawn are, \emph{mutatis mutandis}, analogous to those already reported for Figure \ref{tiresCombined2}, with the unique difference that the estimator predicts quite accurately the slope of the saturated deformations undergone by the front tires. However, this may be easily explained: since the deformation trends depicted in Figure \ref{tiresCombined3} closely resemble the stationary ones, the smart tire sensor mounted on the front axle is essentially measuring the slip angle.

Concerning both the maneuvers investigated in the present Section, some performance indicators of the controller are reported in Table \ref{table:Perf}.
\begin{table}[]
\centering
\begin{tabular}{llll}
\toprule
\textbf{Variable}                     & \textbf{Indicator} & \textbf{Maneuver 1} & \textbf{Maneuver 2} \\
\midrule
\multirow{2}{*}{$\abs{y_{O,\textnormal{ref}}-y_O}$} & RMS (m)               & 0.231               & 0.074               \\
                                      & max (m)                & 1.430             & 0.334               \\
\midrule
\multirow{2}{*}{$\abs{\psi\ped{ref} - \psi}$}  & RMS (-)               & 0.043               & 0.012               \\
                                      & max (-)               & 0.500             & 0.100              \\
\bottomrule
\end{tabular}
\caption{Performance indicators for the two sine maneuvers described in Section \ref{sect:pathfoll}.}\label{table:Perf}
\end{table}

\section{Conclusions}\label{sect:concl}
This paper was concerned with the theoretical development of a model-based, output-feedback lateral tracking control strategy for all-wheel steering vehicles, explicitly incorporating distributed tire dynamics alongside the adoption of intelligent tire sensors. The proposed output-feedback controller was additionally complemented by an observer design strategy aimed at estimating the vehicle’s kinematic variables, tire slip angles, and lateral forces. The estimation problem was addressed under the assumption that smart tire sensors are available in addition to conventional sensor signals. Given the integration of intelligent tire technologies, which necessitates the use of distributed tire representations, the control model employed in this study was formulated as a linear single-track model with distributed tires, represented as a coupled system of ordinary differential equations (ODEs) and partial differential equations (PDEs), where the ODEs describe the rigid-body vehicle dynamics, whilst the PDEs capture the distributed tire behavior.

The effectiveness of the proposed algorithms was tested using a higher-fidelity, nonlinear double-track vehicle model, considering distributed tire dynamics and accounting for finite friction and lateral load transfers. Specifically, two key scenarios were investigated in the validation phase: mitigating micro-shimmy phenomena and achieving accurate path-following via force-based control. Overall, the observer and controller synthesized in this paper demonstrated robust performance, enabling precise estimation of lumped state variables, as well as effective tracking of the desired vehicle’s trajectory in terms of lateral position and yaw.

To the best of the authors’ knowledge, this work constitutes the first rigorous attempt to control vehicular systems equipped with transient, distributed tire representations, in conjunction with smart tire technologies. Future research should focus on developing adaptive observers capable of estimating additional parameters in real-time, such as tire cornering stiffness and relaxation length. Moreover, the proposed control algorithms warrant further validation through experimentation on real vehicles. Finally, the analysis commenced in this paper should be extended to consider scenarios close to the limit of handling, where the tires operate in the nonlinear region, and simultaneously experience high longitudinal and lateral slips. This may be accomplished, for instance, by adopting gain-scheduling control strategies or considering more accurate control-oriented vehicle representations. 

\section*{Acknowledgment}
The authors gratefully acknowledge financial support from the project FASTEST (Reg. no. 2023-06511), funded by the Swedish Research Council.

\section*{Declaration of interest}
Declaration of interest: none.



\begin{thebibliography}{1}

\bibitem{Gerdes1} T. P. Weber, R. K. Aggarwal and J. C. Gerdes, "Human-Inspired Autonomous Racing in Low Friction Environments," IEEE Transactions on Intelligent Vehicles, doi: 10.1109/TIV.2024.3462253.

\bibitem{Gerdes2} T. P. Weber and J. C. Gerdes, "Modeling and Control for Dynamic Drifting Trajectories," in IEEE Transactions on Intelligent Vehicles, vol. 9, no. 2, pp. 3731-3741, doi: 10.1109/TIV.2023.3340918.

\bibitem{Gerdes3} J. Talbot, M. Brown and J. C. Gerdes, "Shared Control Up to the Limits of Vehicle Handling," in IEEE Transactions on Intelligent Vehicles, vol. 9, no. 1, pp. 2977-2987, doi: 10.1109/TIV.2023.3300989.

\bibitem{JazarNew} S. Milani, H. Marzbani and R. N. Jazar, "Vehicle drifting
dynamics: discovery of new equilibria," Vehicle System Dynamics, vol. 60, no. 6, pp. 1933-1958, 2022, doi: \url{https://dx.doi.org/10.1080/00423114.2021.1887499}.

\bibitem{IEEEVT1}
H. Du, N. Zhang and G. Dong, "Stabilizing Vehicle Lateral Dynamics With Considerations of Parameter Uncertainties and Control Saturation Through Robust Yaw Control," IEEE Transactions on Vehicular Technology, vol. 59, no. 5, pp. 2593-2597, Jun 2010, doi: 10.1109/TVT.2010.2045520.

\bibitem{IEEEVT2}
H. Zhou, F. Jia, H. Jing, Z. Liu and L. Güvenç, "Coordinated Longitudinal and Lateral Motion Control for Four Wheel Independent Motor-Drive Electric Vehicle," IEEE Transactions on Vehicular Technology, vol. 67, no. 5, pp. 3782-3790, May 2018, doi: 10.1109/TVT.2018.2816936

\bibitem{Pacejka2} H. B. Pacejka, "Tire and vehicle dynamics," 3rd ed., Amsterdam, Elsevier/BH, 2012.

\bibitem{LibroMio}
L. Romano, "Advanced Brush Tyre Modelling," SpringerBriefs in Applied Sciences. Springer, Cham, 2022. Available from: \url{https://doi.org/10.1007/978-3-030-98435-9}.

\bibitem{CarcassDyn}
L. Romano, F. Bruzelius and B. Jacobson, "Transient tyre models with a flexible carcass," Vehicle System Dynamics, 2023.


\bibitem{Guiggiani}  
M. Guiggiani, "The Science of Vehicle Dynamics," 3rd ed. Springer International, Cham, 2023. Available from: \url{https://doi.org/10.1007/978-3-031-06461-6}.



\bibitem{Meccanica2}
L. Romano, F. Timpone, F. Bruzelius, B. Jacobson, "Analytical results in transient brush tyre models: theory for large camber angles and classic solutions with limited friction," Meccanica 57, 165-191 (2022), doi: \url{http://dx.doi.org/10.1007/s11012-021-01422-3}. 

\bibitem{LuGreSpin}
L. Romano, F. Bruzelius, B. Jacobson, "An extended LuGre-brush tyre model for large camber angles and turning speeds," Vehicle Syst. Dyn. 2022, doi: \url{https://doi.org/10.1080/00423114.2022.2086887}.


\bibitem{Rajamani} R. Rajamani, "Vehicle Dynamics and Control," 1st ed. Springer New York, NY (2012), doi: \url{https://doi.org/10.1007/978-1-4614-1433-9}.

\bibitem{Nielsen} U. Kiencke, L. Nielsen, "Automotive Control Systems," 2nd ed. Springer Berlin, Heidelberg (2005).

\bibitem{Savaresi} S. Savaresi, M. Tanelli, "Active Braking Control Systems Design for Vehicles," Springer, London (2010). 

\bibitem{IEEEVT3}
K. Nam, H. Fujimoto and Y. Hori, "Lateral Stability Control of In-Wheel-Motor-Driven Electric Vehicles Based on Sideslip Angle Estimation Using Lateral Tire Force Sensors," IEEE Transactions on Vehicular Technology, vol. 61, no. 5, pp. 1972-1985, Jun 2012, doi: 10.1109/TVT.2012.2191627.

\bibitem{LuGreControl2} Yu H, Qi Z, Duan J, Taheri S, Ma Y. Multiple model adaptive backstepping control for antilock braking system based on LuGre dynamic tyre model. International Journal of Vehicle Design. 2015;69(1-4):168-184.

\bibitem{RajamaniCC} R. Rajamani, N. Piyabongkarn, J. Lew, K. Yi and G. Phanomchoeng, "Tire-Road Friction-Coefficient Estimation," IEEE Control Systems Magazine, vol. 30, no. 4, pp. 54-69, Aug. 2010, doi: 10.1109/MCS.2010.937006.

\bibitem{Sideslip1} B. C. Chen and F. C. Hsieh, "Sideslip angle estimation using extended Kalman filter," Vehicle System Dynamics, no. 46, sup. 1, pp. 353-364, 2008, doi: \url{https://doi.org/10.1080/00423110801958550}.

\bibitem{Sideslip2} T. Gräber, S. Lupberger, M. Unterreiner and D. Schramm, "A Hybrid Approach to Side-Slip Angle Estimation With Recurrent Neural Networks and Kinematic Vehicle Models," IEEE Transactions on Intelligent Vehicles, vol. 4, no. 1, pp. 39-47, March 2019, doi: 10.1109/TIV.2018.2886687.

\bibitem{Doumiati2} M. Doumiati, A. Victorino, D. Lechner, et al., "Observers for vehicle tyre/road forces estimation: experimental validation," Vehicle Syst. Dyn. vol. 48, no. 11, pp. 1345-1378, 2010.

\bibitem{Hsu} Y. H. J. Hsu, S. M. Laws and J. C. Gerdes, "Estimation of tire slip angle and friction limits using steering torque", IEEE Transactions on Control Systems Technology, vol. 18, no. 4, pp. 896–907, 2010.

\bibitem{Shao1} L. Shao, C. Jin, C. Lex and A. Eichberger, "Nonlinear adaptive observer for side slip angle and road friction estimation," 2016 IEEE 55th Conference on Decision and Control (CDC), pp. 6258-6265, 2016.

\bibitem{Shao3} L. Shao, C. Jin, C. Lex and A. Eichberger, "Robust road friction estimation during vehicle steering," Vehicle Syst. Dyn. vol. 57, no. 4, pp. 493-519, 2019.

\bibitem{AdaptiveCornering} S. You, J. Hahn, H. Lee, "New adaptive approaches to real-time estimation of vehicle sideslip angle," Control Engineering Practice, vol. 17, no. 12, pp. 1367-1379, 2009. Available: \url{https://doi.org/10.1016/j.conengprac.2009.07.002}. 


\bibitem{IntTireSyst} N. Xu, H. Askari and A. Khajepour, "Intelligent Tire Systems," 1st ed., Springer Cham, 2022: DOI: \url{https://doi.org/10.1007/978-3-031-10268-4}.

\bibitem{IT1} S. Yang, Y. Chen, R. Shi, R. Wang, Y. Cao and J. Lu, "A Survey of Intelligent Tires for Tire-Road Interaction Recognition Toward Autonomous Vehicles," IEEE Transactions on Intelligent Vehicles, vol. 7, no. 3, pp. 520-532, Sept. 2022, doi: 10.1109/TIV.2022.3163588.

\bibitem{IT2} H. Lee and S. Taheri, "Intelligent Tires?A Review of Tire Characterization Literature," IEEE Intelligent Transportation Systems Magazine, vol. 9, no. 2, pp. 114-135, Summer 2017, doi: 10.1109/MITS.2017.2666584. 

\bibitem{ITTerrain} S. Khaleghian and S. Taheri, "Terrain classification using intelligent tire," Journal of Terramechanics, vol. 71, pp. 15-24, 2017, doi: \url{https://doi.org/10.1016/j.jterra.2017.01.005}.

\bibitem{ITwear} B. Li, Z. Quan, S. Bei, L. Zhang and H. Mao, "An estimation algorithm for tire wear using intelligent tire concept," Proceedings of the Institution of Mechanical Engineers, Part D: Journal of Automobile Engineering, vol. 235, no. 10-11, pp. 2712-2725, 2021, doi:10.1177/0954407021999483

\bibitem{ITSideSlip} N. Xu, Y. Huang, H. Askari and Z. Tang, "Tire Slip Angle Estimation Based on the Intelligent Tire Technology," IEEE Transactions on Vehicular Technology, vol. 70, no. 3, pp. 2239-2249, March 2021, doi: 10.1109/TVT.2021.3059432.

\bibitem{Benefits} V. Mazzilli, D. Ivone, S. De Pinto, et al., "On the benefit of smart tyre technology on vehicle state estimation," Vehicle System Dynamics, vol. 60, no. 11, pp. 3694-3719, 2021, doi: \url{https://doi.org/10.1080/00423114.2021.1976414}.



\bibitem{Erdogan} G. Erdogan, "New sensors and estimation systems for the measurement of tire-road friction coefficient and tire slip variables," Retrieved from the University Digital Conservancy, 2009. Available from: \url{https://hdl.handle.net/11299/57827}.

\bibitem{IT3} N. Xu, H. Askari, Y. Huang, J. Zhou and A. Khajepour, "Tire Force Estimation in Intelligent Tires Using Machine Learning," IEEE Transactions on Intelligent Transportation Systems, vol. 23, no. 4, pp. 3565-3574, April 2022, doi: 10.1109/TITS.2020.3038155.

\bibitem{ModelBIT1} J. Yi, "A Piezo-Sensor-Based “Smart Tire" System for Mobile Robots and Vehicles," IEEE/ASME Transactions on Mechatronics, vol. 13, no. 1, pp. 95-103, Feb. 2008, doi: 10.1109/TMECH.2007.915064.

\bibitem{ModelBIT3} S. Hong, G. Erdogan, K. Hedrick and F. Borrelli, "Tyre-road friction coefficient estimation based on tyre sensors and lateral tyre deflection: modelling, simulations and experiments," Vehicle System Dynamics, vol. 51, no. 5, pp. 627–647, 2013, doi: \url{https://doi.org/10.1080/00423114.2012.758859}.

\bibitem{ModelBIT2} D. Jeong, S. Kim, J. Lee, S. B. Choi, M. Kim and H. Lee, "Estimation of Tire Load and Vehicle Parameters Using Intelligent Tires Combined With Vehicle Dynamics," IEEE Transactions on Instrumentation and Measurement, vol. 70, pp. 1-12, 2021, Art no. 9502712, doi: 10.1109/TIM.2020.3031124.

\bibitem{Unina1} G. Breglio et al., "Feel-tire Unina: Development and Modeling of a Sensing System for Intelligent Tires," 2019 IEEE 5th International forum on Research and Technology for Society and Industry (RTSI), Florence, Italy, 2019, pp. 453-458, doi: 10.1109/RTSI.2019.8895537.

\bibitem{Unina3} L. Romano, S. Strano and M. Terzo, "A Model-Based Observer for Intelligent Tire Concepts," 2019 IEEE 5th International forum on Research and Technology for Society and Industry (RTSI), Florence, Italy, 2019, pp. 447-452, doi: 10.1109/RTSI.2019.8895599.

\bibitem{Unina2} L. Romano, S. Strano and M, Terzo, "Synthesis and comparative analysis of three model-based observers for normal load and friction estimation in intelligent tyre concepts," Proceedings of the Institution of Mechanical Engineers, Part D: Journal of Automobile Engineering, vol. 235, no. 6, pp. 1629-1642, 2021, doi:10.1177/0954407020975346.

%

\bibitem{Takacs1} D. Takács, G. Orosz, G. Stépán, "Delay effects in shimmy dynamics of wheels with stretched string-like tyres," European Journal of Mechanics - A/Solids, vol. 28, no. 3, pp. 516-525, 2009.

\bibitem{Takacs2}  D. Takács, G. Stépán, "Micro-shimmy of towed structures in experimentally uncharted unstable parameter domain," Vehicle Syst. Dyn. vol. 50, no. 11, pp. 1613-1630, 2012.

\bibitem{Takacs3} D. Takács, G. Stépán, S. J. Hogan, "Isolated large amplitude periodic motions of towed rigid wheels," Nonlinear Dyn. vol. 52, pp. 27–34, 2008, doi: \url{https://doi.org/10.1007/s11071-007-9253-y}.

\bibitem{Takacs4} D. Takács, G. Stépán, "Experiments on quasiperiodic wheel shimmy," ASME. J. Comput. Nonlinear Dynam. vol. 4, no. 3, 2009, doi: \url{https://doi.org/10.1115/1.3124786}.

\bibitem{Takacs5} D. Takács, G. Stépán, "Contact patch memory of tyres leading to lateral vibrations of four-wheeled vehicles," Phil. Trans. R. Soc. A.37120120427, 2013, doi: \url{http://doi.org/10.1098/rsta.2012.0427}.


\bibitem{BicyclePDE} L. Romano, O. M. Aamo, J. Åslund, E. Frisk, "Stability analysis of linear single-track models with transient tyre dynamics", Vehicle System Dynamics, 2024.

\bibitem{ChinaFeng}
L. Zhang, X. Yin, Z. Wang, F. Sun and X. Ding, "Hierarchical Hybrid Steering Control for Four-Wheel-Steering Vehicles Considering System Delays," IEEE Transactions on Vehicular Technology, doi: 10.1109/TVT.2024.3473305.

\bibitem{LarsAVEC}
L. O. Jonsson, A. Balachandran, J. Zhou, B. Olofsson, L. Nielsen, "Investigating Characteristics and Opportunities for Rear-Wheel Steering," 16th International Symposium on Advanced Vehicle Control. AVEC 2024. Lecture Notes in Mechanical Engineering. Springer, Cham, 2024.

\bibitem{Jerrelind}
W. Zhang, L. Drugge, M. Nybacka, J. Jerrelind and Z. Wang, "Exploring Four-Wheel Steering for Trajectory Tracking of Autonomous Vehicles in Critical Conditions," Advances in Dynamics of Vehicles on Roads and Tracks III. IAVSD 2023. Lecture Notes in Mechanical Engineering. Springer, Cham, 2023.

\bibitem{AllWheel1}
T. Kaneko, H. Iizuka and I. Kageyama, "Steering control for advanced guideway bus system with all-wheel steering system," Vehicle System Dynamics, 44, sup. 1, pp. 741–746, 2006. URL: \url{https://doi.org/10.1080/00423110600885731}.

\bibitem{AllWheel2}
F. Fahimi, "Full drive-by-wire dynamic control for four-wheel-steer all-wheel-drive vehicles," Vehicle System Dynamics, vol. 51, no. 3, pp. 360–376, 2012, doi: \url{https://doi.org/10.1080/00423114.2012.743668}.

\bibitem{SteerByWireUlsoy}
A. Kırlı, Y. Chen, C. E. Okwudire and A. G. Ulsoy, "Torque-Vectoring-Based Backup Steering Strategy for Steer-by-Wire Autonomous Vehicles With Vehicle Stability Control," IEEE Transactions on Vehicular Technology, vol. 68, no. 8, pp. 7319-7328, Aug. 2019, doi: 10.1109/TVT.2019.2921016.

\bibitem{Tuononen1} A. J. Tuononen, "Optical position detection to measure tyre carcass deflections," Vehicle System Dynamics, vol. 46, no. 6, pp. 471-481, 2008, doi: \url{https://doi.org/10.1080/00423110701485043}

\bibitem{Tuononen2} A. J. Niskanen and A. J. Tuononen, "Three 3-axis accelerometers fixed inside the tyre for studying contact patch deformations in wet conditions," Vehicle System Dynamics, vol. 52, sup. 1, pp. 287–298, 2014, doi: \url{https://doi.org/10.1080/00423114.2014.898777}.

 \bibitem{MioStand} L. Romano, "Nonlinear modelling of transient tyre dynamics," Nonlinear Dyn, 2025, doi: \url{https://doi.org/10.1007/s11071-025-10997-5}.

\bibitem{Zwarth}
R. Curtain and H. Zwarth, "Introduction to Infinite-Dimensional Systems Theory: A State-Space Approach," 1st ed. Springer New York, NY, 2020, doi: \url{https://doi.org/10.1007/978-1-0716-0590-5}.

\bibitem{CurtainAutomatica}
R. Curtain, K. Morris, "Transfer functions of distributed parameter systems: A tutorial," Automatica, vol. 45, no. 5, pp. 1101-1116, 2009. Available: \url{https://doi.org/10.1016/j.automatica.2009.01.008}.

\bibitem{Pazy}
A. Pazy, "Semigroups of Linear Operators and Applications to Partial Differential Equations," 1st. ed. Springer New York, NY, 1983, doi: \url{https://doi.org/10.1007/978-1-4612-5561-1}.

\bibitem{Tanabe} H. Tanabe, "Functional Analytic Methods for Partial Differential Equations," 1st ed. CRC Press, 1997.

\bibitem{Fors1} V. Fors, B. Olofsson, L. Nielsen, "Formulation and interpretation of optimal braking and steering patterns towards autonomous safety-critical manoeuvres," Vehicle System Dynamics, vol. 57, no. 8, pp. 1206-1223, 2019, doi: \url{https://doi.org/10.1080/00423114.2018.1549331}.

\bibitem{Fors2} V. Fors, B. Olofsson, L. Nielsen, "Attainable force volumes of optimal autonomous at-the-limit vehicle manoeuvres," Vehicle System Dynamics, vol. 58, no. 7, pp. 1101-1122, 2020, doi: \url{https://doi.org/10.1080/00423114.2019.1608363}.

\bibitem{Fors3} V. Fors, B. Olofsson, L. Nielsen, "Autonomous Wary Collision Avoidance," IEEE Transactions on Intelligent Vehicles, vol. 6, no. 2, pp. 353-365, 2021, doi: \url{https://doi.org/10.1109/TIV.2020.3029853}.

\bibitem{Fors4} V. Fors, P. Anistratov, B. Olofsson, L. Nielsen, "Predictive Force-Centric Emergency Collision Avoidance," ASME. J. Dyn. Sys., Meas., Control. vol. 143, no. 8, pp. 081005, 2021, doi: \url{https://doi.org/10.1115/1.4050403}.










\end{thebibliography}
\end{document}